\documentclass[10pt,journal,final,onecolumn]{IEEEtran}
\usepackage[utf8]{inputenc}
\usepackage{amsmath}
\usepackage{amsfonts}
\usepackage{amssymb}
\usepackage{graphicx}
\usepackage{float}
\allowdisplaybreaks
\usepackage{multirow,array}
\usepackage{mathrsfs}
\usepackage{float}
\usepackage{comment}

\usepackage{amsmath}
\usepackage{amsfonts}
\usepackage{amssymb}
\usepackage{graphicx}
\usepackage{float}
\usepackage{longtable}
\usepackage{multirow,array}
\usepackage{float}

\usepackage{fancyhdr}
\usepackage{bm}
\usepackage{amsmath}

\usepackage{booktabs}
\usepackage{adjustbox}
\usepackage{longtable}
\usepackage{longtable,lscape}
\usepackage{rotating}
\usepackage[graphicx]{realboxes}
 
\usepackage{xcolor}

\usepackage{fancyhdr}
\usepackage{amsmath}
\usepackage{amsfonts}
\usepackage{amssymb}
\usepackage{graphicx}
\usepackage{float}

\usepackage{multirow,array}
\usepackage{float}
\usepackage{fancyhdr}
\usepackage{bm}
\usepackage{amsmath}

\usepackage{booktabs}
\usepackage{adjustbox}
\usepackage{longtable}
\usepackage{longtable,lscape}
\usepackage{rotating}
\usepackage[graphicx]{realboxes}
\usepackage{amsmath}

\usepackage{booktabs}
\usepackage{adjustbox}
\usepackage{longtable}
\usepackage{longtable,lscape}
\usepackage{rotating}
\usepackage[graphicx]{realboxes}
\begin{document}
	
	
	%
	\setlength\parindent{2em}

	\title{ A joint latent class model of longitudinal and survival data with  a time-varying membership probability }
	\author{Ruoyu Miao; Christiana Charalambous \\
		The University of Manchester
	}
	\date{2020/12/13}  
	\maketitle  
	\begin{abstract}
		\quad Joint latent class modelling has been developed considerably in the past two decades. In some instances, the models are linked by the latent class $k$ (i.e. the number of subgroups), in others they are joined by shared random effects or a heterogeneous random covariance matrix.  We propose an extension to the joint latent class model (JLCM) in which probabilities of subjects being in latent class $k$ can be set to vary with time. This can be a more flexible way to analyse the effect of treatments to patients. For example, a patient may be in period \textrm{I} at the first visit time and may move to period \textrm{II} at the second visit time, implying the treatment the patient had before might be noneffective at the following visit time. For a dataset with these particular features, the joint latent class model which allows jumps among different subgroups can potentially provide more information as well as more accurate estimation and prediction results compared to the basic JLCM. A Bayesian approach is used to do the estimation and a DIC criterion is used to decide the optimal number of classes. Simulation results indicate that the proposed model produces accurate results and the time-varying JLCM outperforms the basic JLCM. We also illustrate the performance of our proposed JLCM on the aids data (Goldman et al., 1996).\\
		
	\end{abstract}
	
	\section{Introduction}
		\indent \quad Joint models for longitudinal and survival data are a popular approach for handling the dependencies between these two types of data. They are particularly useful for modelling longitudinal data with informative dropout or performing time to event analysis with time dependent covariates, but they can also be used if the interest is on both processes, while also considering their association. Incorporating all information can help increase the efficiency and decrease the bias of inferences (Ibrahim et al., 2010). \\
			\indent \quad Joint modelling of longitudinal measurements and survival data was first introduced in the early 90s. Schluchter (1992) pointed out the problem of non-ignorable censoring data in longitudinal studies and gave an approach which was based on lognormal survival. Self and Pawitan (1992), De Gruttola and Tu (1994) and Tsiatis et al. (1995) applied joint modelling approaches to AIDS research, which adjusted the inferences on the longitudinal responses with informative missing data by proposing the two-stage method to do the parameter estimation. Henderson et al. (2000) proposed a flexible joint model, where the longitudinal and survival processes were linked via a latent Gaussian process, allowing for both serial correlation and measurement error in the longitudinal model. Rizopoulos (2009) proposed a new computational approach, in the form of a fully exponential Laplace approximation for the joint modelling of survival and longitudinal data, making is feasible to handle high dimensional random effects structures in joint models.  Motivated by a primary biliary cirrhosis (PBC) study (Murtaugh et al., 1994), a joint modelling of multi-state event times and longitudinal data with informative observation time points was proposed by Dai  and Pan (2018), where  the joint modelling was linked by random effects under no distributional assumption.	 This model  extended  the 	corrected score method (Wang, 2006) to cases with longitudinal data which is collected at informative time points.\\
				\indent \quad
			Typically, joint models are linked by shared parameters/random effects or by random effects sharing the same distribution. The development of specialised software, such as JM (Rizopoulos, 2010), joineR (Philipson et al., 2012), stjm (Crowther et al., 2013) and JMBayes (Rizopoulos, 2016) among others, has also helped popularise this type of approach. Other work in the literature includes copula joint models (Kürüm et al., 2018, Zhang et al., 2021) as well as hidden Markov models (Bartolucci  and  Farcomeni, 2019; Zhou et al., 2021). Further extensions to consider multiple longitudinal outcomes and/or competing risks (Elashoff et al.,  2008; Hu et al., 2009;  Huang et al., 2011; 	Proust-Lima et al., 2015; Rouanet, A. et al., 2016) have also been considered. \\
			\indent \quad  Traditional joint modelling via shared random effects in a homogenous group, though widely used,  is limited to describe only one pattern for subjects.  In many  circumstances, the presence of heterogeneous subgroups can not be ignored and thus  potential subtypes of  longitudinal and survival outcomes should be considered. In order to deal with the underlying subpopulations with different patterns of  responses, joint latent class models can be used. These models are  usually applied to analyse data from health and aging studies, e.g. prostate cancer, AIDS or dementia studies.  Lin et al. (2002) first proposed the joint latent class model (JLCM)  by extracting meaningful subgroups with respect to the joint distribution of the longitudinal data and the event process, which could be used to model informative dropout and help  identify patterns of  longitudinal measurements over time.  They proposed and applied a joint latent class model of longitudinal PSA and the PCA onset, permitting distinct behaviours within each subgroup to model the association between  longitudinal and  time-to-event data. Garre et al. (2008) proposed a joint latent class changepoint model to improve the prediction of time to graft failure. They divided  patients into two different groups, such that subjects in latent class 1 were modelled by an  intercept-only random-effects model and subjects within  latent class 2 were modelled by a segmented random changepoint model, achieving better predictions of time to graft failure compared to  other joint models.\\
				\indent \quad 	Chiang (2009) postulated a more flexible joint latent varying-coefficient model to handle  population heterogeneity, which generalised  the correlation mechanism in the joint  model with two latent processes, proposed by Henderson et al.  (2000). Additionally, Liu et al. (2015) introduced a joint latent class model by integrating the joint random effects model and applying this model to an AIDS study to analyse the relationship between longitudinal CD4+ measurements and the time to death. This type of proposed model is appropriate when nonignorable heterogeneity exists  among subjects. An overview  of earlier work on   joint latent class models is provided  in a review article by Proust-Lima et al. (2014). Proust-Lima et al. (2017) also developed the R package lcmm  which can implement various extended mixed models, including  joint latent class models, using the maximum likelihood method.    More recently, Andrinopoulou et al. (2020) introduced a Bayesian approach to joint latent class models, which incorporated a new latent class selection procedure via a mixture model and applied the proposed approach to cystic fybrosis data.\\
	\indent \quad 	Reboussin and Anthony (2001) proposed the  dynamic  latent class regression model for longitudinal data by adding   time-varying covariates into the baseline-category logistic regression model,  Lin et al. (2014) added random effects to the aforementioned   dynamic latent class model for longitudinal data, to  jointly model  the informative event and  a class-specific logistic model with shared random effects. Zhang and Simonoff (2020)  discussed both the strengths and weaknesses of the joint latent class modelling  and presented an idea for a  nonparametric joint modelling approach. They pointed that  the joint latent class tree (JLCT) model  with a tree-based approach could be a good alternative by addressing  the time-invariant limitation of the JLCM.  Zhang and Simonoff (2022) proposed a tree-based approach to model time-to-event and longitudinal data. This semiparametric joint latent class tree model uses time varying covariates as the splitting variables for constructing a tree in the first stage. Once the latent class has been identified for each individual at each time-point, in the second stage, it can fit the longitudinal and survival models independently.  Despite the advantages in prediction performance and computational cost, the JLCT can not handle the scenario where longitudinal and survival outcomes are associated even after conditioning on the latent class.\\
	\indent \quad   Motivated by Garre et al. (2008) and Liu et al. (2015),  we  want to explore changes in time in subpopulations of  subjects. 	Under this situation, patients  may experience pathological changes such that a particular treatment regime might stop being effective for a patient and a different regime might need to be considered. This could be reflected by a change in the latent class for that patient. In order to handle this kind of particular dataset with jumping behaviours, we propose adding time-varying covariates into the latent class membership probability, which allows jumps among different subgroups. \\ 
		\indent \quad  Based on the standard JLCM, the  JLCM  with a  time-varying membership probability is proposed, where the longitudinal and survival processes are joint via shared random effects as well as  the latent class $k$. The proposed time-varying JLCM can relax  the time-invariant restrictions of JLCM and also capture jumping behaviors for longitudinal markers. In contrast to the dynamic latent class model proposed by Reboussin and Anthony (2001), we consider a multinomial logistic regression  latent class submodel. Compared with the Bayesian JLCM proposed by Andrinopoulou  et al. (2020), we compute dynamic membership probabilities  for each subject $i$ across time, by introducing time-varying covariates in the latent class submodel.\\
			\indent \quad  	 A Bayesian approach is used for estimation and inference. The detailed structure of the time-varying  JLCM  is presented in section \textrm{II}. In section \textrm{III}, we conduct a simulation study to compare the performance of the proposed JLCM with the basic JLCM when there is jumping behaviour in the data, while in section  \textrm{IV}, we apply our model to analyse the AIDS dataset (Goldman et al., 1996).   For both simulation and real data analyses, we also investigate the performance of the proposed model in the dynamic prediction of survival probabilities. Finally, section  \textrm{V} summarises the work in this paper and discusses future extensions of the proposed model. \\
	
	\section{Joint latent class model with time-varying probability}
		\indent\quad  Joint latent class modelling has been developed greatly in the past two decades. However, basic JLCMs can only allow one pattern for classification. In some instances, the class may change with time for some subjects which may experience different pathological changes.   For example, in the CPCRA Aids study (Abrans et al., 1994; Neaton et al., 1994), CD4 counts can be  divided into several latent classes clearly without mixing up.  Different CD4 counts stand for different health status, and treatments regimes can be changed for patients with time. Individual may be in high CD4 level (good health status) at the first visit time and may move to low CD4 level (bad health status) at the second visit time, implying   any previous treatments might be noneffective at the following visit time. Under this situation, the potential pathological changes cannot be ignored. In order to deal with these underlying changes in treatments regimes, the time-varying JLCM is proposed by adding time-varying covariates into the latent class submodel. Without considering the jumping behaviors, treatments regimes could stop being effective for patients.\\
	\indent\quad We define our  reference basic JLCM is consist of the latent class submodel without time-varying covariates, the longitudinal and survival submodels which have shared random effects. Based on the basic JLCM, the structure of our proposed JLCM is shown as follows.
	
	\subsection{General latent class submodel with time varying probability}
	\quad We consider a logistic model to describe the class membership probability, allowing this probability to vary with time $t$. Suppose there are $N$ subjects, labelled  $i=1,2, \dots N$ and  $K$ latent classes labelled  $k=1,2, \dots K$.    Assume the design matrices $\bm{X}_{1i}$ may have time-dependent covariates or polynomial terms in time  measured at time points $t_i=t_{ij}|_{j=1,2, \dots m_i}= (t_{i1}, t_{i2}, \cdots, t_{im_i})^T$, where $t_{ij}$ denotes the $j$th visit time of subject $i$ ($j = 1, 2, \dots  m_i$) and $m_i$ is the
number of obervations for subject $i$. 	Define $R_{ij}=k$ as the latent class indicator, which means subject $i$ belongs to class $k$ at time point $t_{ij}$. The probability $\pi_{ijk}$ for  subject $i$ measured at time $t_{ij}$, of belonging to  class $k$ with $\sum_{k=1}^{K}\pi_{ijk}=1$ is described using a multinomial logistic regression model with covariate vector $\bm{X}_{1i}$ and associated class-specific coefficient vector $\bm{\xi}_{k}$, as follows:\\ 
	$$\begin{aligned}
		\pi_{ijk}=P(R_{ij}=k)=\frac{\exp(\bm{X}_{1i}^T(t_{ij})\bm{\xi}_{k})}{\sum\limits_{s=1}^{K}\exp(\bm{X}_{1i}^T(t_{ij})\bm{\xi}_{s})} \  \  \ \forall~~k=1, \ldots, K
	\end{aligned}  
	\eqno{(1)}
	$$
	
	\subsection{Longitudinal submodel}
	\indent\quad   Denoting $y_{ijk}$ as a repeated measurement for subject $i$ at the $jth$ visit in  class $k$,  the longitudinal submodel for $y_{ij}$ is (Troxel et al., 1998):
	$$\begin{aligned}
		y_{ij}|(R_{ij}=k)=\bm{X}_{2i}^T(t_{ij})\bm{\beta}_{k}+\bm{Z}_{i}^T(t_{ij})\bm{U}_{ik}+\epsilon_{ijk}
		\ \ \ \ \ K \ne 1
	\end{aligned}  
	$$
	where $\bm{X}_{2i}(t_{ij})$ is the covariate vector of subject $i$, which might contain time-varying covariates or time polynomial terms,  $\bm{\beta}_{k}$ is the parameter vector for class $k$ and $\bm{U}_{ik}$ ($q \times 1$) are the class-specific random effects for class $k$. $\bm{Z}_{i}(t_{ij})$ denotes the vector of covariates  at the $jth$ visit time associated with  $\bm{U}_{ik}$, again allowing for time-varying covariates or time polynomial terms. $\epsilon_{ijk} \stackrel{i.i.d} \sim N(0,\tau_k)$ is the class-specific error term with  variance  $\tau_k$. We assume $\epsilon_{ijk}$ is independent of $\bm{U}_{ik}$ (Liu et al., 2015). The class-specific random effects $\bm{U}_{ik}$ can also be written as $\bm{U}_{ik}=\mathfrak{Z}_k\bm{U}_i$, where $\bm{U}_i \sim N_{q}(0, \Sigma_{u_i})$. For simplification, we assume $\mathfrak{Z}_k=1$, i.e.  we assume the random effects across different latent classes are the same. Then, a simplified form of the longitudinal submodel is:
	$$\begin{aligned}
		y_{ij}|(R_{ij}=k)=\bm{X}_{2i}^T(t_{ij})\bm{\beta}_{k}+\bm{Z}_{i}^T(t_{ij})\bm{U}_{i}+\epsilon_{ijk}
		\ \ \ \ \ K \ne 1
	\end{aligned}  
	\eqno{(2)}
	$$
	\subsection{Survival submodel}
	\indent \quad Let $T^*_i$ denote the true event time which may or may not be observed ($i=1,2, \dots N$). Let $C_i$ denote the censoring time for subject $i$. The observed follow-up time is represented by $T_i=min(T_i^*, C_i)$. Define $\ddot{\Delta}_i=I(T_i^* \le C_i)$ as the censoring indicator, where $\ddot{\Delta}_i$ takes the value $0$ if  censoring occurs and $1$ otherwise. The time to event in each latent class can be described by a Cox frailty model with class-specific baseline hazard function and parameters (Hougaard, 1995):
	$$\begin{aligned}
		\lambda_{i}(t|R_{ij}=k)=\lambda_{0k}(t)\exp(\bm{X}_{3i}^T(t)\bm{\omega}_{k}+\bm{Z}_{i}^T(t_{})(\bm{\delta}_{k}\bm{U}_i))
	\end{aligned}  
	\eqno{(3)}
	$$
	where $\lambda_{0k}(t)$ denotes the baseline hazard function in class $k$, $\bm{X}_{3i}(t)$ is the covariate vector for subject $i$ at time $t$ and  $\bm{\omega}_{k}$ is the parameter vector for class $k$. In addition,  $\bm{\delta}_{k}$ denotes the  association vector for class $k$ associated with the random effects $\bm{U}_i$, which is considered as the strength of the association between the longitudinal measurements and survival outcomes. If $\bm{\delta}_{k}=\bm{0}$, it means the longitudinal   and survival submodels are only linked via the latent class $k$.\\
	\indent \quad We further assume that the $kth$ baseline hazard is a step function as follows
	$$\begin{aligned}
		\lambda_{0k}(t)=\lambda_{0k}^{(s)},  ~~~~~~ t_{k}^{(s-1)}<t \le t_{k}^{(s)}
	\end{aligned}  
	\eqno{}
	$$
	where $0<t_{k}^{(1)}< \dots <t_{k}^{(S^k)}< \infty$ is a partition of $(0, \infty)$, $k=1,2, \dots K$ and $S_k$ denotes the number of steps for the $k$th baseline hazard, labelled as $s = 1, 2, \dots S^k$.
	
	\subsection{Estimation}
	
	\subsubsection{The likelihood }
	
	\indent \quad We  denote $\bm{\Psi}$ as the  collection of model parameters,   such that $\bm{\Psi}=(\bm{\xi}_{}, \bm{\beta}_{}, \bm{\omega}_{},  \bm{\delta}_{},   \bm{\tau}_{},  \bm{\lambda}_{},  \bm{\Sigma}_{u})$ where $\bm{\xi}=(\bm{\xi}_{1}, \dots, \bm{\xi}_{K})$, $ \bm{\beta}=(\bm{\beta}_{1}, \dots, \bm{\beta}_{K})$, $\bm{\omega}=(\bm{\omega}_{1}, \dots, \bm{\omega}_{K})$, $\bm{\delta}=(\bm{\delta}_{1}, \dots, \bm{\delta}_{K})$, $\bm{\tau}=({\tau}_{1}, \dots, {\tau}_K)$, $\bm{\lambda}=({\lambda}_{01}, \dots, \lambda_{0K})$, and $\bm{R}_{i}=(R_{i1}, \ldots, R_{im_i})$. Assuming that the observed data $(\bm{Y},\bm{T},\bm{\ddot{\Delta}})$ are independent conditional on the latent class indicator $\bm{R}$ and the random effects $\bm{U}$, and that $(\bm{T},\bm{\ddot{\Delta}})$ only depends on $R_{im_i}$, the likelihood contribution for the $i$th patient is written as
$$\begin{aligned}
	L_i(\bm{\Psi}|\bm{R}_{i}=k,\bm{Y}_{i},T_i,\ddot{\Delta}_i,\bm{U}_i)
	=f(\bm{Y}_{i}|\bm{R}_{i}=k,\bm{U}_i,\bm{\Psi})f(T_i|\ddot{\Delta}_i, {R}_{im_i}=k,\bm{U}_i,\bm{\Psi}) P(\bm{R}_{i}=k|\bm{\Psi}) f(\bm{U}_i|\bm{\Psi}) 
\end{aligned}  
$$
Then, we obtain the joint likelihood function \\
$$\begin{aligned}
	L_{}(\bm{\Psi}|\bm{R}, \bm{Y}, \bm{T},\bm{\ddot{\Delta}},\bm{U})
	=\prod_{i=1}^{N}f(\bm{U_i}|\bm{\Psi})\prod_{j=1}^{m_i}\prod_{k=1}^{K}P_{ijk} 
\end{aligned}  
\eqno{(4)}
$$
where 
\begin{align*}
	P_{ijk}&=\left\{
	\begin{array}{ll}
		\{P(R_{ij}=k|\bm{\Psi})f(y_{ij}|R_{ij}=k, \bm{U}_i,\bm{\Psi})\}^{I(R_{ij}=k)}
		&
		\quad\text{for  } j \neq m_i \\
		\\
		\{P(R_{im_i}=k|\bm{\Psi})f(y_{im_i}|R_{im_i}=k, \bm{U}_i,\bm{\Psi})f(T_i|\ddot{\Delta}_i, {R}_{im_i}=k,\bm{U}_i,\bm{\Psi})\}^{I(R_{im_i}=k)}
		&
		\quad\text{for  } j = m_i ,
	\end{array}
	\right. 
\end{align*} \\
$$f(y_{ij}|(R_{ij}=k, \bm{U}_i,\bm{\Psi})= (2\pi\tau_k )^{-{1}/{2}}\exp\{-{1}/({2\tau_k})(y_{ij}- \bm{X}_{2i}^T(t_{ij})\bm{\beta}_{k}-\bm{Z}_{i}^T(t_{ij})\bm{U}_i)^2\},$$ 
$$f(T_i|\ddot{\Delta}_i, {R}_{im_i}=k,\bm{U}_i,\bm{\Psi}) = \lambda_{ik}(t|R_{im_i}=k, \bm{U}_i,\bm{\Psi})^{\ddot{\Delta}_i} \exp\{-H_{ik}(t_{}|R_{im_i}=k, \bm{U}_i,\bm{\Psi})\}$$
and 
$$f(\bm{U}_i|\bm{\Psi})=(2\pi)^{-{q}/{2}}det(\Sigma_{ui})^{-{1}/{2}}\exp(-{tr(\Sigma_{ui}^{-1}\bm{U}_i\bm{U}_i^T)}/{2}).$$
\setcounter{equation}{4}
In addition, 
$$\lambda_{ik}(t|R_{ij}=k, \bm{U}_i,\bm{\Psi})^{\ddot{\Delta}_i}=
(\lambda_{0k}(t)\exp(\bm{X}_{3i}^T(t)\bm{\omega}_{k}+\bm{Z}_{i}^T(t_{})(\bm{\delta}_{k}\bm{U}_i)))^{I(T_i^*\le C_i)},$$ 
\begin{align}
	H_{ik}(t_{}|R_{ij}=k, \bm{U}_i,\bm{\Psi})&=\int_{0}^{t_{}}
	\lambda_{0k}(u)\exp(\bm{X}_{3i}^T(u)\bm{\omega}_{k}+\bm{Z}_{i}^T(u_{})(\bm{\delta}_{k}\bm{U}_i))du \nonumber\\   
	&=\sum_{s=1}^{S^k}I(t_{}\ge t_k^{s-1})\lambda_{0k}^{(s)}\int_{t_k^{(s-1)}}^{min(t_{},t_{k}^{(s)})}
	\exp(\bm{X}_{3i}^T(u)\bm{\omega}_{k}+\bm{Z}_{i}^T(u_{})(\bm{\delta}_{k}\bm{U}_i))du.
\end{align}
	
		\subsubsection{MCMC sampling procedure}
	\indent \quad We employ a Bayesian estimation approach to obtain the parameter estimates in our proposed JLCM. Based on the likelihood given in (4), the posterior distribution is given by:
	$\pi(\bm{\Psi}|\bm{R}, \bm{Y}, \bm{T},\bm{\ddot{\Delta}},\bm{U}) \propto L(\bm{\Psi}|\bm{R}, \bm{Y}, \bm{T},\bm{\ddot{\Delta}},\bm{U}) \pi(\bm{\Psi})$ 
	where $\pi(\bm{\Psi})$ is the prior distribution of parameter vector $\bm{\Psi}$. Assuming apriori independence, we can write $\pi(\bm{\Psi})$ as the product of the prior for each parameter component.
	We can draw $\bm{\beta}_{},\bm{\tau}_{},\bm{\lambda}_{}, \Sigma_{ui}$ directly from their full conditional distributions using Gibbs sampling. We set  normal priors for $\bm{\beta}_{}, \bm{\omega}_{}, \bm{\delta}_{}, \bm{U}_{}, \bm{\xi}_{}$, obtaining  conjugate posterior distributions for $\bm{\beta}_{}$. We also choose a gamma distribution as the prior for $\bm{\lambda}_{}$, an inverse Wishart prior for $\Sigma_{ui}$ and an inverse gamma distribution as the prior for $\bm{\tau}_{}$. Details of the calculation of the full conditional densities for $\bm{\beta}_{}$, $\bm{\tau}_{}$, $\bm{\lambda}_{}$ and $\Sigma_{ui}$, can be found in  Appendix A. We also sample $\bm{R}_{ij}$ directly from its discrete posterior distribution where the posterior probability of subject $i$ at time $j$  being a member of latent class $k$ is $P(R_{ij}=k \ | \bm{R}, \bm{Y}, \bm{T},\bm{\ddot{\Delta}},\bm{U})={P_{ijk}}/({P_{ij1}+ \cdots +P_{ijK}})$. \\
	\indent \quad However, for $\bm{w}_{}, \bm{\delta}_{}, \bm{\xi}_{}, \bm{U}$ the conditional distributions have non-standard densities, making sampling from these distributions more complex.  A traditional  Metropolis-Hastings type algorithm (Metropolis et al., 1953; Hastings 1970) although widely used and easy to implement, can often have convergence issues. Adaptive MCMC algorithms can help to deal with this problem by automatically learning better parameter values of MCMC algorithms while they run (Roberts \& Rosenthal, 2009). Hence, an adaptive MCMC method is preferred to  sample the parameters/components with non-standard distributions in our proposed JLCM.
	Due to convergence issues with sampling $\bm{\tau}$ and $\bm{\lambda}$ directly from their conditional posteriors, we  sample them via the AM scheme, instead.\\
	\indent \quad According to the Adaptive Metropolis (AM) algorithm in Haario et al.   (2001), we consider a $\ddot{d}$-dimensional target distribution $\pi(\bm{\theta}, \bm{\tau}, \bm{\lambda})$, where $\bm{\theta}=(\bm{w}, \bm{\delta}, \bm{\xi}, \bm{U} )$ and define  $\bm{\theta}_p=(\bm{w}, \bm{\delta}, \bm{\xi})$, with $\ddot{d}_p$ denoting the dimension of $\bm{\theta}_p$. $\ddot{d}_{\tau}$ is defined as the dimension of $\bm{\tau}$ and $\ddot{d}_{\lambda}$ is defined as the dimension of $\bm{\lambda}$. We assume that the elements of $\bm{\theta}_p$ have individual independent standard normal  priors   $N(0,1)$ with the density function denoted as $f_{N(0,1)}$ and the elements of vector  $\bm{\tau}$ have individual independent vague inverse Gamma  priors $IG(0.01, 0.01)$ with the density function denoted as $f_{IG(0.01, 0.01)}$. We also set that  the elements of vector  $\bm{\lambda}$ to have individual independent uninformative  Gamma  priors $\Gamma(0.01, 0.01)$ with the density function denoted as $f_{\Gamma(0.01, 0.01)}$.
		$$\begin{aligned}
		\pi(\bm{\theta},\bm{\tau}, \bm{\lambda}|\bm{R}, \bm{Y}, \bm{T},\bm{\ddot{\Delta}},\bm{\Psi}_{-(\bm{\theta}_p,\bm{\tau}, \bm{\lambda})})
		&\propto\Big\{\prod_{i=1}^{N}f(\bm{U}_i|\bm{\Psi}_{-(\bm{\theta}_p,\bm{\tau}, \bm{\lambda})}\prod_{j=1}^{m_i}\prod_{k=1}^{K}P_{ijk} \Big\} \\
		& \prod\nolimits_{ \ddot{d}_p}f_{N(0,1)}\prod\nolimits_{ \ddot{d}_{\tau}}f_{IG(0.01, 0.01)}\prod\nolimits_{ \ddot{d}_{\lambda}}f_{\Gamma(0.01, 0.01)}\\
	\end{aligned}  
	\eqno{(6)}
	$$
where $\bm{\Psi}_{-(\bm{\theta}_p,\bm{\tau},\bm{\lambda})}$ is the parameter vector excluding the components in $\bm{\theta}_p$, $\bm{\tau}$ and $\bm{\lambda}$.  Then, we can constuct an adaptive Metropolis algorithm with proposal distribution $\pi(\bm{\theta}, \bm{\tau},\bm{\lambda}|\bm{R}, \bm{Y}, \bm{T},\bm{\ddot{\Delta}}, \bm{\Psi}_{-(\bm{\theta}_p,\bm{\tau},\bm{\lambda})})$ given at iteration $m$ by
	
	$$\begin{aligned}
		Q_m( (\bm{\theta}, \bm{\tau},\bm{\lambda}),\cdotp)=\left\{
		\begin{aligned}
			&N( (\bm{\theta}, \bm{\tau},\bm{\lambda}), (0.1)^2I_{\ddot{d}}/\ddot{d} ) \ \ \  \ \ \ \ \ \ \ \ \ \ \ \mbox{for}\  m \leq 2\ddot{d} \\
			&(1-\alpha_{prop})N( (\bm{\theta}, \bm{\tau},\bm{\lambda}), \sigma^2\cdotp \Sigma_m/\ddot{d})\\
			&+ \alpha_{prop}\cdotp N( (\bm{\theta}, \bm{\tau},\bm{\lambda}), (0.1)^2I_{\ddot{d}}/\ddot{d} ) \ \ \  \mbox{for}\  m>2\ddot{d} \\
		\end{aligned}
		\right.
	\end{aligned}  
	\eqno{(7)}
	$$
	where $\Sigma_m$ denotes the current empirical estimate of the covariance structure of the target distribution based on the runs so far, $\sigma^2$ is  a fixed constant variance  which we choose, to determine how well the adaptive algorithm is performing  and ${\alpha}_{prop}$ is a small positive constant. 
	Gelman et al. (1997) and Roberts and Rosenthal (2001) noted that the proposal $N(\bm{\theta}, (2.38)^2\Sigma_m/\ddot{d})$ is optimal in a particular large-dimensional context. Here, we found that $N( \bm{\theta}, (0.1)^2I_{\ddot{d}}/\ddot{d} )$ is a more suitable proposal.  \\
	\indent\quad According to the usual  Metropolis-Hastings formula (Hastings, 1970), a proposal from  $(\bm{\theta}, \bm{\tau},\bm{\lambda})$  to  $ (\bm{\theta}^*, \bm{\tau}^*,\bm{\lambda}^*)$ is accepted with   probability  $$\alpha((\bm{\theta}, \bm{\tau},\bm{\lambda}), (\bm{\theta}^*, \bm{\tau}^*,\bm{\lambda}^*))=min(\frac{{\pi((\bm{\theta}^*, \bm{\tau}^*,\bm{\lambda}^*)}|\bm{R}, \bm{Y}, \bm{T},\bm{\ddot{\Delta}}, \bm{\Psi}_{-(\bm{\theta}_p,\bm{\tau},\bm{\lambda})})}{\pi((\bm{\theta}, \bm{\tau},\bm{\lambda})|\bm{R}, \bm{Y}, \bm{T},  \bm{\ddot{\Delta}}, \bm{\Psi}_{-(\bm{\theta}_p,\bm{\tau},\bm{\lambda})})}, 1)$$
	where $\bm{\theta}^*=(\bm{w}^*, \bm{\delta}^*, \bm{\xi}^*, \bm{U}^*)$ and  
	$\bm{\theta}^*$,  $\bm{\tau}^*$ and $\bm{\lambda}^*$ are vectors of  parameters for $\bm{\tau}$ and $\bm{\lambda}$, which are the generated ``candidate values" from $Q_m( (\bm{\theta}, \bm{\tau},\bm{\lambda}) , \cdotp)$.\\
	\indent \quad Through the MCMC sampling scheme, we  can also calculate the  posterior membership probability to  perform model-based classification. The posterior probability of subject $i$ belonging to class $k$ at time $j$ is
	\begin{align*}
		\hat{\pi}_{ijk} (\hat{\bm\Psi})=\frac{	\pi_{ijk}L_{ijk}(\hat{\bm\Psi})}{\sum\limits_{k=1}^{K}	\pi_{ijk}L_{ijk}(\hat{\bm\Psi})} 
	\end{align*}  
	$L_{ijk}(\hat{\bm\Psi})=f(\hat{\bm U}_i|\hat{\bm\Psi})\hat{P}_{ijk}$ is computed using the posterior sample of the parameters $\bm{\Psi}$ and random effects $\bm{U}$, and $\pi_{ijk}$ is the prior probability from equation (1).  The greater the value of the posterior probability $\hat{\pi}_{ijk}  (\hat{\bm\Psi})$ is, the more likely the subject is divided into this group.\\ 
	
		\subsubsection{Misclassification rate}	
	\indent \quad   The misclassification rate, also known as the error rate,  is calculated as the percentage of miss-classified counts over the whole dataset. For example, if we only have two classes, then the error rate for the JLCM without time-varying probability is computed as  (Garre et al., 2008):
	$$\begin{aligned}
		Error \  rate =\frac{n_{M1}+n_{M2}}{N}
	\end{aligned}
	$$
	where $n_{M1}$ and  $n_{M2}$ denote  the counts which are mis-classified into class 1 and class 2, respectively and $N$ denotes the total number of subjects.\\
	\indent \quad For the time-varying JLCM, we proposed a  similar measure to calculte the error rate. 
	$$\begin{aligned}
		Error \  rate =\frac{N_{M1}+N_{M2}}{\sum\limits_{i=1}^{N}m_i}
	\end{aligned}
	\eqno{(8)}
	$$
	where $N_{M1}$ and  $N_{M2}$ denote  the counts which are mis-classified into class 1 and class 2 across time for all individuals and  $m_i$ denotes the number of observations for subject $i$.\\
	
		\subsubsection{Predicted time to event}	
	\indent \quad  Following Garre (2008) and  Rizopoulos (2011), we can compute  dynamic predictions of the time-to-event probabilities based on our proposed JLCM. The survival function for  subject $i$ is given by 
	$$\begin{aligned}
		S_i(t+\Delta t|T\geq t, y_i(t), \hat{\bm{\Psi} }, \hat{\bm{U} }_i )=\frac{ S_i\{ t+\Delta t| y_i(t), \hat{\bm{\Psi} }, \hat{\bm{U} }_i  \}  }{   S_i\{ t| y_i(t), \hat{\bm{\Psi} }, \hat{\bm{U} }_i  \}   }
	\end{aligned}
	\eqno{(9)}
	$$
	where $ \hat{\bm{\Psi} }$ and $\hat{{U} }_i $ denote the posterior means of the parameter vector $ {\bm{\Psi} }$ and random effects $U_i$, and  $y_i(t)$ contains the longitudinal measurements  up  to time $t$  to imply  survival up to this time point. $	\hat{S}_i\{ t| y_i(t),R_{ij}=k,\hat{\bm{\Psi} }, \hat{\bm{U} }_i  \}=\exp(-	\hat{H}_{ik}(t| y_i(t),R_{ij}=k, \hat{\bm{\Psi} }, \hat{\bm{U} }_i    ))$ with $\hat{H}_{ik}(\cdot)$ given in (5). 
	The posterior predictive survival function $S_i$ at $t$ to predict the survival of a patient up to time $t$ is given by
	$$\begin{aligned}
		\hat{S}_i\{ t|, y_i(t), \hat{\bm{\Psi} }, \hat{\bm{U} }_i \}=\sum_{k=1}^{K} \hat P(R_{im_i}=k| \hat{\bm{\Psi} })	\hat{S}_i(t| y_i(t),R_{im_i}=k,\hat{\bm{\Psi} }, \hat{\bm{U} }_i)
	\end{aligned}
	\eqno{(10)}
	$$	
	which is a  mixture  of survival distributions for patient $i$ conditional on subgroup $k$ to get  survival probabilities for each subject.\\
	\indent \quad 	To assess the prediction accuracy of the proposed model, we calculate the IPCW estimators of the   area under the ROC curve (AUC) within the time interval $[t, t+\Delta t)$  for joint models of longitudinal markers and time-to-event outcomes (Blanche et al., 2015) as  follows:
		$$\begin{aligned}
			\widehat{AUC}{(t, \Delta t)}=\frac{\sum\limits_{i=1}^{N}\sum\limits_{j=1}^{N}	I(	\pi_i(t,  \Delta t)>	\pi_j(t,  \Delta t))	\widehat{D}_i(t, \Delta t)(1-\widehat{D}_j(t, \Delta t)	\widehat{W}_i(t, \Delta t)\widehat{W}_j(t, \Delta t)}{\sum\limits_{i=1}^{N}\sum\limits_{j=1}^{N}\widehat{D}_i(t, \Delta t)(1-\widehat{D}_j(t, \Delta t)	\widehat{W}_i(t, \Delta t)\widehat{W}_j(t, \Delta t)}
		\end{aligned}
		\eqno{(11)}
		$$	
		where
		$$\begin{aligned}
			\widehat{D}_i(t, \Delta t)=I(t  \leq T_i < t+\Delta t)
		\end{aligned}
		$$	
		$$\begin{aligned}
			\widehat{W}_i(t, \Delta t)=\frac{I(T_i \geq t+\Delta t)}{\widehat{G}(t+\Delta t|t)}+\frac{I(t  \leq T_i < t+\Delta t)\ddot{\Delta}_i}{\widehat{G}(T_i|t)}
		\end{aligned}
		$$	
		$	\widehat{D}_i(t, \Delta t)$ is an  indicator denoting 1 if subject $i$ experienced the event within $[t, t+\Delta t)$ and  0 otherwise. $\widehat{G}(u)$ is the Kaplan-Meier estimator of the censoring time at $u$ in the   survival function and $\widehat{G}(u|s)=\widehat{G}(u)/\widehat{G}(s)$ for $ {\forall} u>s$ to estimate the  conditional probability of being uncensored at time point $u$, conditionally on not being censored at time point $s$ (Blanche et al., 2015). Notations of $T_i$ and $\ddot{\Delta}_i$ can be found in the descriptions of the   survival submodel and  $\pi_i(t,  \Delta t)$ is given   in equation (9).\\

	\section{Simulation study for joint latent class model with a time-varying probability }

	\subsection{Simulation setup}

	\indent \quad 
 For simplicity,   in our simulation  study we just consider two latent classes
	and  use baseline covariates in  $X_{3i}$  (fixed effects in the survival model) but a more complex model with time-varying $X_{3i}$  is also possible. The fixed  covariate $X_{1i}$ which is generated from a standard normal distribution, is used to capture certain  baseline characteristics of each subject and $X_{3i}$  corresponds to a  treatment indicator simulated from the bernoulli distribution  (1 if subject receives treatment  and 0 otherwise). The time variable $Time_{ij}$ can take values 0, 0.05, 0.1, 0.15, 0.2, 0.25, 0.3, 0.35, 0.4, 0.45, 0.5, 0.55, 0.6, 0.65, 0.7, 0.75,
	0.8, 0.85 and each subject has at least 6 measurements. 
	The non-informative censoring time is sampled from an exponential distribution with mean 6. The longitudinal submodel is in the form  $y_{ijk}=\beta_{k1}X_{1i}+\beta_{k2}Time_{ij}+U_{1i}+U_{2i}Time_{ij}+\epsilon_{ijk} $ and the survival submodel is $\lambda_{ijk}=\lambda_{0k}\exp(w_kX_{3i}+\delta_{k1}U_{1i}+\delta_{k2}U_{2i}Time_{ij})$ where  $\bm{U}_i \sim N(\bm{0},\bm{I}_2)$. where  $\bm{U}_i \sim N(\bm{0},\bm{I}_2)$.
	Here we set  a constant baseline hazard function $h_{0k} (t)=\lambda_{0k} $ (i.e. we sample the  baseline hazard  $\lambda_{0k} $ from an exponential distribution). We simulate survival times using  $T_k=log[1-({U_{2} log(u)})/({\lambda_{0k} \exp(w_kX_{3}+U_{1})})]/{U_{2}} $ where $u\sim U(0,1)$. For the membership probability, we  set $p_{ijk}={\exp(\xi_{k1}X_{1i}+\xi_{k2}Time_{ij})}/({\sum_{c=1}^K(\exp(\xi_{c1}X_{1i}+\xi_{c2}Time_{ij}))})$.  $p_{ijk}$ is used to classify subgroups  based on a  multinomial distribution, which is known in advance.  The error rate can be calculated by the percentage of the counts of being mis-classified into class 1 and class 2 for the whole dataset (equation (8)).\\
	\indent\quad Each dataset is generated as follows:\\
	$$\bm{U}_i \sim N_{(q)} \begin{pmatrix} 
		\begin{pmatrix} 0\\0 \end{pmatrix},\Sigma_{ui} =\begin{pmatrix}
			1 \qquad 0\\
			0 \qquad 1
	\end{pmatrix} \end{pmatrix} $$ \\
	\textbf{Class 1}:\\
	Latent class: $p_{ij1}=\frac{\exp(0.01X_{1i}+0.2Time_{ij})}{(\exp(0.01X_{1i}+0.2Time_{ij})+\exp(Time_{ij}))}$ \\
	Longitudinal submodel: $y_{ij1}=2X_{1i}+0.5Time_{ij}+U_{1i}+U_{2i}Time_{ij}+\epsilon_{ij1} $\\ 
	We specify  $\epsilon_{ij1} \stackrel{i.i.d} \sim N(0,{0.1})$ \\
	Survival submodels: $\lambda_{i1}(t)=0.2\exp(0.5X_{3i}-0.5U_{1i}-0.8U_{2i}t_{}) \  \ \ for \ t \le T_i$\\
	
	\textbf{Class 2}:\\
	Latent class: $p_{ij2}=\frac{\exp(Time_{ij})}{\exp(0.01X_{1i}+0.2Time_{ij})+(\exp(Time_{ij}))}$ \\
	Longitudinal submodel: $y_{ij2}=4X_{1i}+3Time_{ij}+U_{1i}+U_{2i}Time_{ij}+\epsilon_{ij2} $\\ 
	We specify $\epsilon_{ij2} \stackrel{i.i.d} \sim N(0,{0.5})$ \\
	Survival submodels: $\lambda_{i2}(t)=0.1\exp(0.8X_{3i}-1.5U_{1i}-0.4U_{2i}t_{})  \  \ \ for \ t \le T_i$\\
	
	\subsection{Simulation results}
	\quad For  each simulation run, we set $N$ as the total number of subjects and $m_i$ as the number of repeated longitudinal measurements for each subject $i$. Here is an example for one of 100 simulations with $5000$ MCMC iterations. We have $N=50$  individuals with at least 6 repeated measurements for each subject $i$.   The sample size for this particular dataset is $\sum\limits_{i=1}^{N}m_i=620$ , with observations  at time 0, 0.05, 0.1, 0.15, 0.2, 0.25, 0.3, 0.35, 0.4, 0.45, 0.5, 0.55, 0.6, 0.65, 0.7, 0.75, 0.8, 0.85. The dataset is divided into 2 classes. The censoring rate is   80.32\% (0: censored data because of drop off; 1: the event happens because of death). Among the 50 subjects, 47  have  jumps between the two subgroups.  
	\begin{figure}[H]
		\begin{center}
				\includegraphics[width=1.05\textwidth, height=0.37\textwidth]{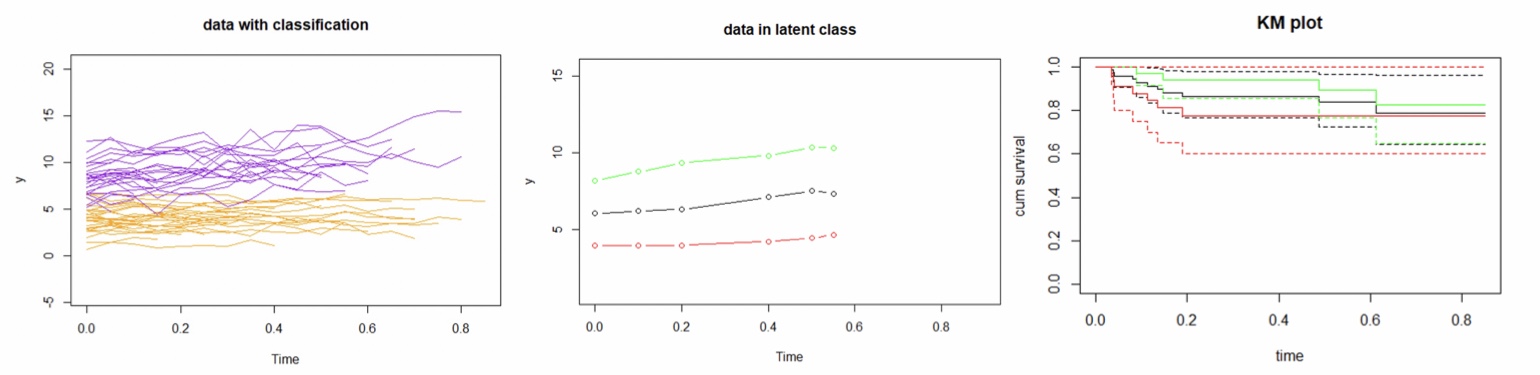}
			\center{Fig. 1.   Left: Longitudinal trajectories with certain classification (orange: class 1/low level; purple: class 2/high level); Middle: Mean longitudinal trajectories; Right: K-M plot (black: no classification; red: class 1/low level; green:
				class 2/high level)}
		\end{center}
	\end{figure}

	\indent \quad Firstly we focus on the basic JLCM which does not allow the group membership probability to change with time.	The plot of longitudinal trajectories is shown in the left plot of Figure 1. 
The plot of longitudinal trajectories classisfied without time-varying changes is presented in the middle plot  of Figure 1.\\
		\indent \quad 	The mean longitudinal trace and KM plot based on the basic JLCM without time-varying changes are also shown in Figure 1. It is obvious that allowing for classification is important, to distinguish between the two different subgroups in this dataset.  However, this kind of static classification can not tell us how individuals might change group membership over time. Based on our model, we select eight individuals to show how changes in the group membership are reflected based on the longitudinal trajectories. 
	

	\indent \quad  We generate some new data under the  same settings of simulation study as we have described in subsection B to illustrate group jumping behaviors and do the predictions for some specific subjects.  From Figures 2 and 3, we observe  three different jumping behaviours: some subjects, e.g. 28, 34 and 43 remain stable in the same group (no jumping). Others, e.g. 3 and 38, jump from one group to the other and remain there until the event or censoring occurs. Finally, other subjects, such as 7, 9 and 13 experience multiple jumps between groups. Unless we allow for the group membership probability to vary with time, we cannot capture these different behaviours across individuals. For example, 	for patients undergoing cancer staging (i.e.  the process of determining the extent to which a cancer has developed by spreading), different medicines and treatment measures should be considered  for patients with  moderate  or severe symptoms. Therefore, it becomes important to monitor any changes in the spread of cancer during therapy and adjust treatment plans accordingly. Our proposed model can provide this flexibility by allowing changes in group membership with time.\\
		\begin{figure}[H]
		\begin{center}
			\includegraphics[scale=0.55]{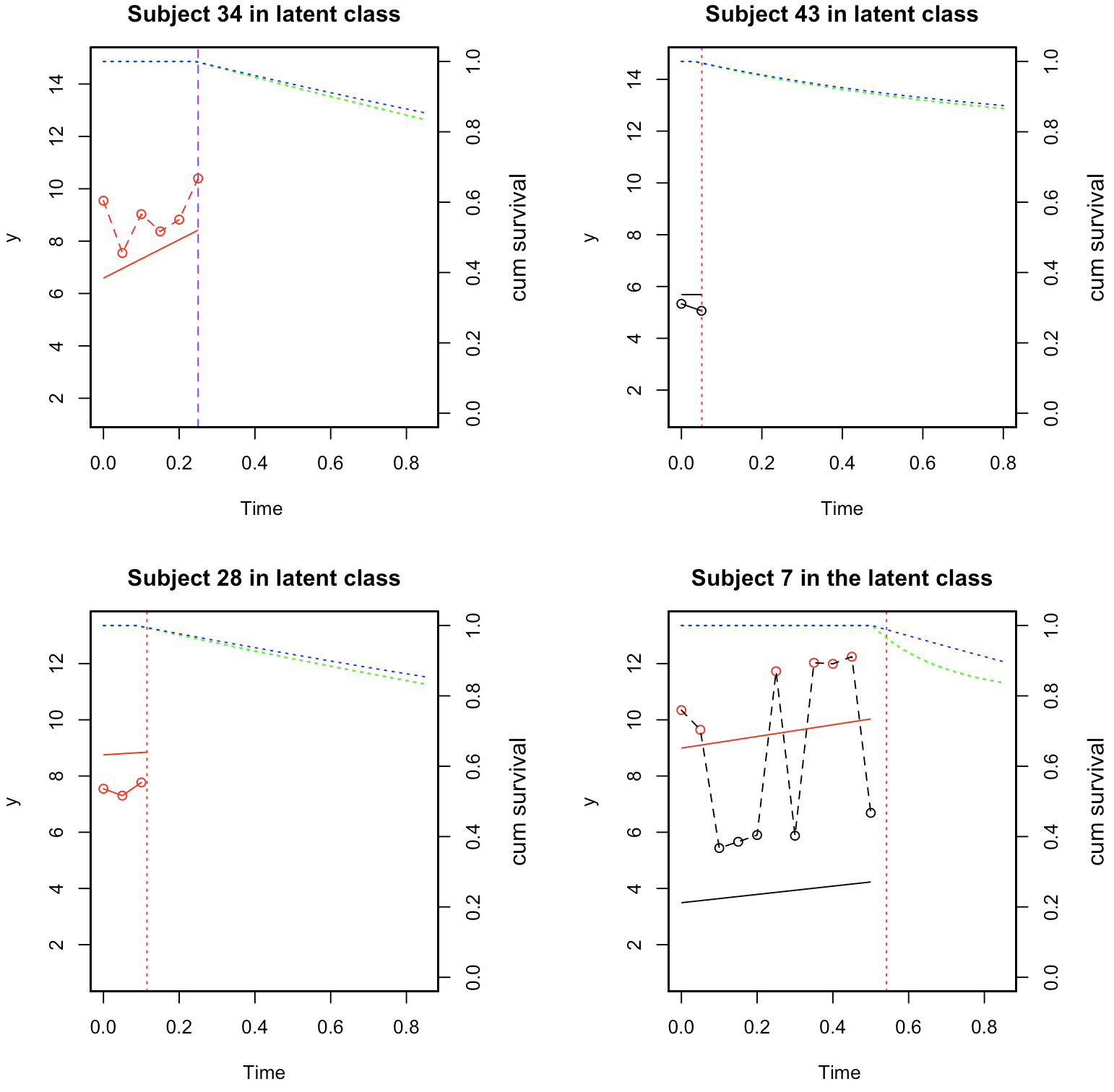}
		\end{center}
		\center{ Fig. 2 Survival predictions of simulated data for four cases from both proposed and basic JLCM (black point: class 1/low level; red point: class 2/high level; black dashed  line: longitudinal trajectory which stayed in class 1; red  dashed  line: longitudinal trajectory which stayed in class 2; black full line: fitted trajectory which stayed in class 1; red  full line: fitted trajectory which stayed in class 2;  green dotted line: survival probability from proposed JLCM; blue dotted line: survival probability from basic JLCM; purple vertical dashed line: subject censored; red vertical  dotted line: time to event happened)
		}
	\end{figure}
	\indent\quad Tables 1  and 2 summarise the results of our simulation study, over 100 simulated datasets and using 5000 MCMC iterations with 2000 iterations as burn-in. The optimal number of classes for each model will be chosen via DIC. From Table 1, we can see an overwhelming support for the correct model, which is the proposed JLCM with time-varying probability and 2 classes, according to both DIC and the error rate. In the small number of cases when the basic JLCM is  selected as the optimal model (only according to DIC), we notice that a model with 3 classes is preferred. This is to be expected, as the basic JLCM might be trying to compensate for any jumping behaviours by introducing an additional class. \\

{ \centerline{ Table 1:	{ DIC and error rate over 100 simulated datasets and 3000 MCMC iterations}
}}
	\begin{longtable}{llllllccc}
		\hline			
		\multicolumn{6}{c}{improved JLCM 	 }	 
		&\multicolumn{1}{c}{basic JLCM} 
		\\ 
			\hline
		Model      & $K$=1   & $K$=2 & $K$=3  \  & $K$=1 & $K$=2 & $K$=3 & $K$=4\\ 	\hline	
		Model selection via DIC  &  \ \ 0& 0.92&  \ \ 0  \ \ \  \   &\ \ 0 &  0.03 & 0.05&0    \\
			Model selection via error rate   & \ \ 0  & \ \ 1  & \ \ 0  \ \ \  \  & \ \ 0  &\ \ 0   &  0 &0 \\
		\hline	
	\end{longtable}




\begin{figure}[H]
	\begin{center}
		\includegraphics[scale=0.7]{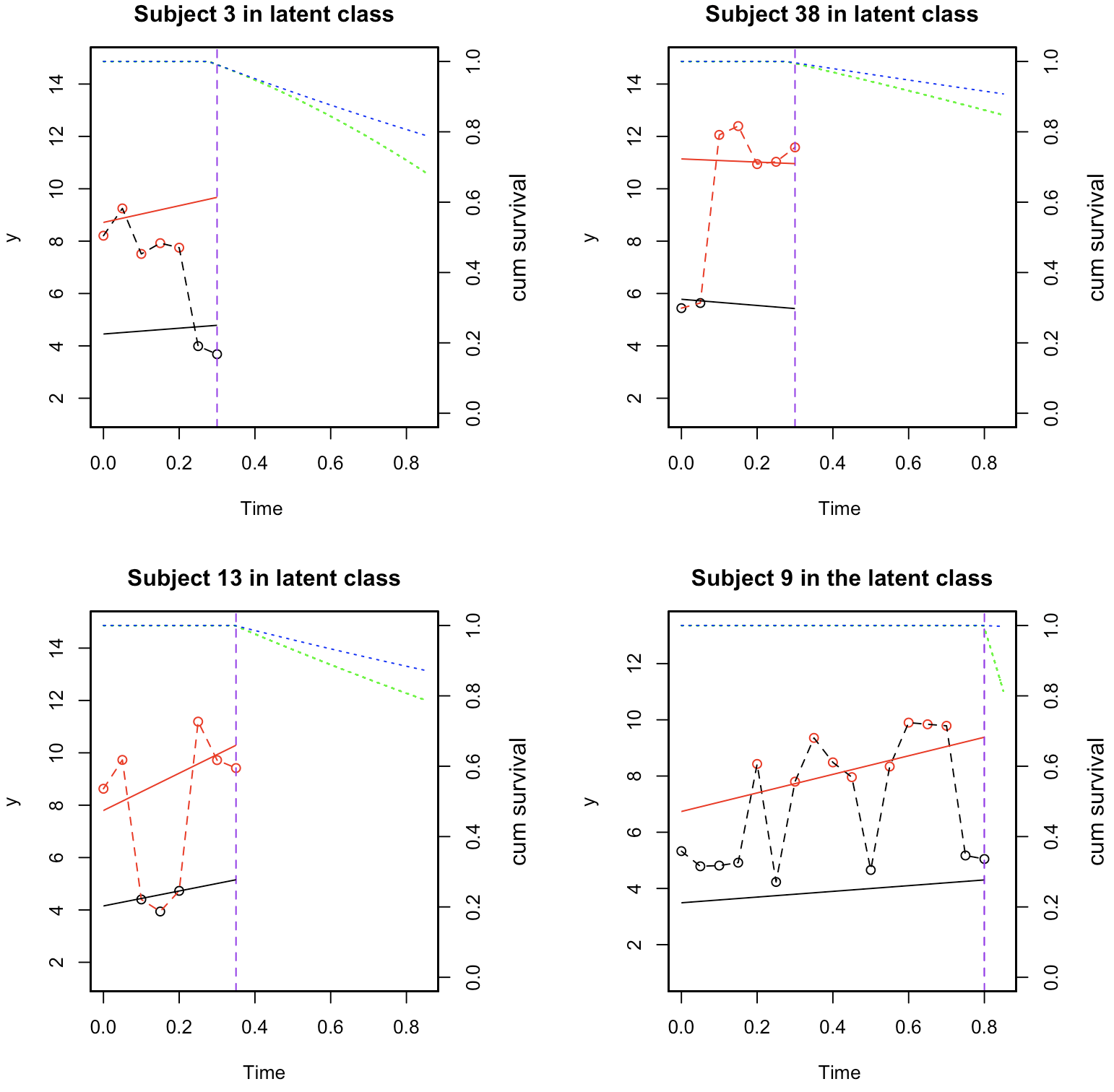}
	\end{center}
	\center{ Fig. 3 Survival predictions of simulated data for four cases from both proposed and basic JLCM (black point: class 1/low level; red point: class 2/high level; black dashed  line: longitudinal trajectory which stayed in class 1; red  dashed  line: longitudinal trajectory which stayed in class 2; black full line: fitted trajectory which stayed in class 1; red  full line: fitted trajectory which stayed in class 2;  green dotted line: survival probability from proposed JLCM; blue dotted line: survival probability from basic JLCM; purple vertical dashed line: subject censored)
	}
\end{figure}	

	\indent\quad  Table 2 indicates low biases for parameter estimates, which further supports the good performance of the proposed JLCM.  Kruschke (2014) suggested that $95\%$ might not be the most appropriate for Bayesian posterior distributions  and McElreath (2018) considered $89$ as the highest prime number which does not exceed the  unstable $95\%$ threshold (Makowski et al., 2019). Here we choose the $89\%$  Credible Interval (CI) with the Equal-tailed Interval (ETI) of posterior distributions using the quantiles method.\\

	{	\centerline{ Table 2: 	{	Estimation results over 100 simulated datasets and 3000 MCMC iterations, based on the time varying JLCM with $K$=2
		}
	}
	}

\setlength{\tabcolsep}{3.3mm}{
	\begin{longtable}{llllllcc}
		\hline
		\multicolumn{6}{c}{\ \ \ \ \ \ \ \ \ \ \ \ \ \ \ \ \ \ \ \ \ \ \ \ \ \ \ \ \ \ \ \ \ \ \ \ \ \ \ \ \ \  time-varying JLCM with 	$K$=2 	 }	 \\ \hline 
			\multicolumn{4}{c}{Class 1}	 
		&\multicolumn{4}{c}{Class 2} 
		\\ 
		\hline
		Parameter           & Estimate (Bias) &sd  & CI(89\% ETI)& 	Parameter  &Estimate (Bias) &sd  & CI(89\% ETI)\\ \hline 
		
		$\beta_{11}$        &  2.0897(0.0897)  & 0.1860&[2.03, 2.25] &	$\beta_{21}$ & 3.9623(-0.0377)   &  0.1284&[3.90, 4.03]\\
		$\beta_{12}$         &   0.5990(0.0990) &  0.7609&[0.35, 0.72]&$\beta_{22}$ & 2.9819(-0.0181)& 0.3078&[2.63, 3.26]\\
		$\tau_1$          & 0.1529(0.0529)  & 0.0849&[0.09, 0.26]&$\tau_2$          & 0.4795(-0.0205) &0.2054 &[0.30, 0.72]\\
		$\lambda_{1}$   &  0.2037(0.0037) &  0.1024&[0.12, 0.37]&$\lambda_2$       & 0.1040(0.0040) & 0.0631&[0.04, 0.19]\\
		$	w_1    $                        & 0.4756(-0.0244)  &  0.2006&[0.31, 0.69]&	$w_2  $                          & 0.7626(-0.0374)  &   0.3096& [0.50, 1.02]\\	
		$\delta_{11}$   &-0.4799(0.0201)  &  0.2106&  [-0.78, -0.32]&	$\delta_{21} $  &-0.7618(0.0382) &  0.3100& [-1.13, -0.51]\\	
		$\delta_{12} $  & -1.4324(0.0676) & 0.5652 &[-2.04, -0.99]&	$\delta_{22}$  & -0.3798(0.0202)   &   0.1812& [-0.64, -0.19]\\	
		$\xi_{11}$           &0.0196(0.0096) &  0.0681& [-0.05, 0.11] &$\xi_{21}$              & -0.0038(-0.0038) & 0.0700&[-0.09, 0.09]\\ 
		$\xi_{12}$           & 0.1891(-0.0109) & 0.1169&[0.08, 0.29]&$\xi_{22}$           & 0.9501(-0.0499) & 0.3628&[0.66, 1.45]\\	
		\hline
	\end{longtable} }

				\subsection{Dynamic predictions of time to death}

			\indent \quad 	We further demonstrate the performance of our proposed model by comparing the expected survival of individuals, to that obtained via the basic JLCM. The survival  predictions are computed using equations (9) and (10) described in Section \textrm{II}, and the results for eight specific subjects are presented in Figures 2 and 3.  
		The plots show the combined information on the longitudinal trajectories and the survival probabilities, for each of the chosen subjects. The  green dotted line is the survival  prediction for proposed JLCM with 2 classifications (the true model), the blue dotted line corresponds to the survival  probabilities of the basic  JLCM with  3  subgroups (the  best  alternative model) and the vertical dashed line indicates the time to event (red) or censoring (purple).\\
		\indent \quad For subject 7 in  Figure 2, we can see that at the time of event, the probability of survival from the proposed model is lower when the event occurred (0.9677501 
		as opposed  0.9894454 
		computed via the basic JLCM). This individual  exhibits jumping behaviours which implies the proposed  model can predict the survival probability for this individual more accurately. For individuals who have a stable trajectory (i.e. subjects 28, 34 and 43), the proposed model provides  marginally lower but  similar survival probabilities to the basic JLCM, which indicates that our proposed JLCM includes the basic model  as a more general JLCM. 
	 For individuals who exhibit jumping behaviours as in  Figure 3, we can see that our proposed model is more conservative and provides lower predictive survival probabilities. It is also more sensitive to changes in group membership,  e.g.  for subject 3, the probability is more reactive to the drop in the response at the final two measurements. The dynamic prediction plots for these  subjects are shown in figure C1 of Appendix C. \\
		\indent \quad
		To compare the prediction accuracy of the proposed and basic JLCM, we calculate the IPCW estimator of the  AUC  within the  time interval $[0.5, 0.8)$ at $t=5$ with $\Delta t=0.3$ for 100 simulations. The boxplots of the IPCW estimator of the  AUC (i.e. $\widehat{AUC}{(0.5, 0.8)}$),  shown in Figure 4, further verify that allowing for the latent class membership probability to change over time can offer improved predictions of survival probabilities.

	\section{Real data application}
	
	\subsection{Data information}
	\indent\quad   The Aids dataset (Goldman et al., 1996) contains both longitudinal and survival data  from a randomized clinical trial looking to compare the efficacy and safety of two antiretroviral drugs in treating patients who had failed or were intolerant of zidovudine (AZT) therapy.  The aids drug trial data is a data frame with 1408 observations on  9 different variables, i.e. $patient$, $Time$, $death$, $CD4$, $obstime$, $drug$, $gender$, $prevOI$ and $AZT$. There  are 467 patients  in total, distinguished by patients identifiers. Time denotes the time to death or censoring. $Death$ is the indicator of a numeric vector with 0 denoting censoring and 1 death. The $CD4$ value is the CD4 cell counts   in blood, which was recorded at study entry in 0, 2, 6, 12 and 18 months. CD4 cells are white blood cells which can fight infection as well as  used to check the health of the immune system in patients infected with human immunodeficiency virus (HIV). $Obstime$
	denotes the time points at which the CD4 cell counts was recorded. $Dru$g is a factor with levels ddC denoting zalcitabine and ddI denoting didanosine. $Gender$ is a factor with levels female and male. $prevOI$ is a factor with levels AIDS, denoting previous opportunistic infection (AIDS diagnosis) at study entry, and noAIDS, denoting no previous infection. $AZT$ is a factor with levels intolerance and failure denoting AZT intolerance and AZT failure, respectively.\\
	
	\begin{figure}[H]
		\begin{center}
			\includegraphics[width=0.9\textwidth, height=0.65\textwidth]{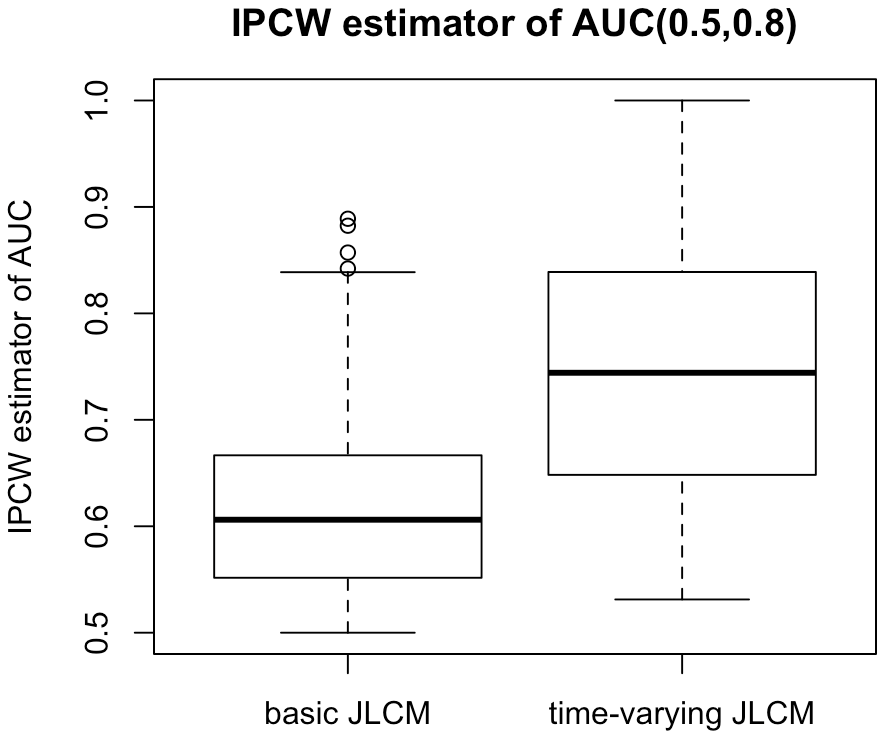}
		\end{center}
		\center{ Fig. 4 Boxplots of the IPCW estimator of AUC when assuming the time-varying and basic JLCM at time point  $t=0.5$ with $\Delta t=0.3$)
		}
	\end{figure}
	
	\indent\quad We analyse the dataset with the basic joint model and our proposed JLCM with time varying probability. 
 Based on the proposed  time-varying JLCM with $K$ subgroups, the longitudinal submodel  is as follows:\\
	$$\begin{aligned}
		CD4_{ijk}=&\beta_{k1}+\beta_{k2}Time_{ij}+\beta_{k3}gender_{ij}+\beta_{k4}prevOI_{ij}+\beta_{k5}AZT_{ij}
		+U_{1i}+U_{2i}Time_{ij}+\epsilon_{ijk}\\
	\end{aligned}
	$$\\
	 the class specific hazards submodel is:\\
	$$\begin{aligned}
		\lambda_{ik}(t)=\lambda_{0k}\exp(w_{k1}gender_i+w_{k2}prevOI_i+w_{k3}AZT_i+w_{k4}drug_i+\delta_{k1}U_{1i}+\delta_{k2}U_{2i}t_{}) \  \ \ \mbox{for} \ t \le T_i\\
	\end{aligned}
	$$\\
	 and the fitted latent class submodel is:\\
	$$\begin{aligned}
		p_{ijk}=\frac{\exp(\xi_{k1}+\xi_{k2}Time_{ij})}{\sum\limits_{s=1}^{2}\exp(\xi_{s1}+\xi_{s2}Time_{ij})}\  \  \ \forall~~k=1, \ldots, K
	\end{aligned}
	$$\\
	\indent \quad	
	$U_{1i}$ is the random intercept and $U_{2i}$ is treated as the random slope for time in the linear mixed-effects model,
	with a bivariate normal distribution $	\bm{U_{i}} \sim N_{(q)}  (  \bm{0},\Sigma_{ui}   )$			
	and $\epsilon_{ijk}$ are the measurement errors which follow a normal distribution: $\epsilon_{ijk} \stackrel{i.i.d} \sim N(0,\tau_k^2)$. \\	
	\indent \quad 
	For comparison, we obtain the  basic JLCM with MLE approach from  ``lcmm" package as following.\\
	
	\indent \quad 
	The longitudinal submodel  is:\\
	$$\begin{aligned}
		CD4_{ijk}=&\beta_{k1}+\beta_{k2}Time_{ij}+\beta_{k3}gender_{ij}+\beta_{k4}prevOI_{ij}+\beta_{k5}AZT_{ij}
		+U_{1i}+U_{2i}Time_{ij}+\epsilon_{ijk}\\
	\end{aligned}
	$$\\
	where  $\epsilon_{ij} \stackrel{i.i.d} \sim N(0,\tau)$ and the class-specific hazard submodel is:\\
	$$\begin{aligned}
		\lambda_{ik}(t)=\lambda_{k}^*\nu_k^*t^{\nu^*_k-1}\exp(w_{1}gender_i+w_{2}prevOI_i+w_{3}AZT_i+w_{4}drug_i) \  \ \ \mbox{for}\ t \le T_i\\
	\end{aligned}
	$$\\
	where the class-specific baseline hazrad is sample from Weinull distribution and the scale parameter $\lambda_{k}^* >0$, the shape parameter $\nu_k^* >0$.\\
	\indent \quad 
	For identification, the last latent class  is treated as reference, i.e.: $\bm{\xi}_{K}=\bm{0}$. The fitted latent class submodel is:\\
	$$\begin{aligned}
		p_{ik}=\frac{\exp(\xi_{k1})}{1+\sum\limits_{s=1}^{K-1}\exp(\xi_{s1})}\  \  \ \forall~~k=1, \ldots, K
	\end{aligned}
	$$\\
	$U_{1i}$ is the random intercept and $U_{2i}$ is treated as the random slope for time in the linear mixed-effects model,
	with a bivariate normal distribution $	\bm{U_{i}} \sim N_{(2)}  (  \bm{0},\Sigma_{ui}   )$			
	and $\epsilon_{ijk}$ are the measurement errors which follow a normal distribution: $\epsilon_{ijk} \stackrel{i.i.d} \sim N(0,\tau_k)$. \\
	
	\subsection{Estimation results and  comparison with results from basic JLCM}

	\indent \quad For the real data application, we consider seven different model settings, similar to our earlier simulation study, including the proposed JLCM with 1, 2 or 3 classes and the basic JLCM with 1-4 classes.\\
	
	{ \centerline	{Table 3: 	{DIC table of aids data for both improved and basic JLCMs in Bayeian approach}
		}
	}
	
		\begin{longtable}{llllllccc}
			\hline			
			\multicolumn{5}{c}{improved JLCM 	 }	 
			&\multicolumn{2}{c}{basic JLCM} 
			\\ 	\hline
			Model      & $K$=1   & $K$=2 & $K$=3 \ \ \ \ \ \ \ \  \ \ \  \  \   & $K$=1 & $K$=2 & $K$=3 & $K$=4\\ 	\hline	
			DIC   & 1999.237 \ \ \  &  1649.333& 2598.579  \  \   \ \ \ \ \    &  12832&  9558.951&   9469.819&   10137.06 \\
			\hline			
		\end{longtable}	
	
		\indent \quad According to Table 3, the proposed JLCM with 2 classes is the optimal model. In terms of the basic JLCM, the optimal model is the one with 3 classes (which matches our findings in the simulation study). This result also agrees with the results obtained via the lcmm package, which fits the basic JLCM under a maximum likelihood approach. More details can be found in Appendix B.\\	
		\indent \quad 	Table 4 summarises the jumping behaviours of individuals in the AIDS dataset. We can see  that most patients remain stable in the same class and only 16\% of patients have jumping behaviours. For patients with jumping behaviours,  around 61\%  jump from the better group (higher CD4 cell count) to the lower CD4 group, which might indicate worsening of their health.  \\
			
	{ \centerline{Table 4: 	{ Jumping behaviours  in the aids dataset based on the   time-varying JLCM with 	$K$=2 using a Bayesian approach}
	}
}
	\begin{longtable}{llllllccc}
		\hline			
		\multicolumn{3}{c}{no moves 	 }	 
		&\multicolumn{2}{c}{jumping behaviours	} 
		\\ 	\hline
		Classification     & class 1   & class 2 \ \ \ \   & in class 1 at the final time point  & in class 2 at the final time point  
		\\ 	\hline	
		Frequency    & \  \  \ 76  \ & \ \ 317 \ \  \  \ \  \   & \ \ \ \ \ \  \ \ \ \ \ \ \ \ \ \ \  29 &  \ \ \ \ \ \ \ \ \ \ \ \ \ \ \ \ 45 
		\\
		\hline
		\multicolumn{1}{c}{		total 	 }	 
		&\multicolumn{2}{c}{ \ \ \ \ \ \ \ \ \ 393 \ \  \ \ \ \ \ \ \  \ \  \	} 
		&\multicolumn{2}{c}{ \ \ \ \ \ \ 74
		}  	\\
		\hline			
	\end{longtable}

			\indent \quad Figure 5 illustrates the differences in the mean trajectories and survival probabilities in the aids data, between a scenario with no classification and one with a static 2 group classification. We notice two distinct groups, with patients with a higher CD4 cell count having no risk of death, compared to those with lower CD4 cell count. The wide confidence bounds around the survival probabilities of the latter group, indicate that perhaps not all patients in this group are associated with lower survival probabilities and allowing for jumps between groups could provide a more accurate prediction of these probabilities.\\
			
				\begin{figure}[H]
					\begin{center}
						\includegraphics[scale=0.35]{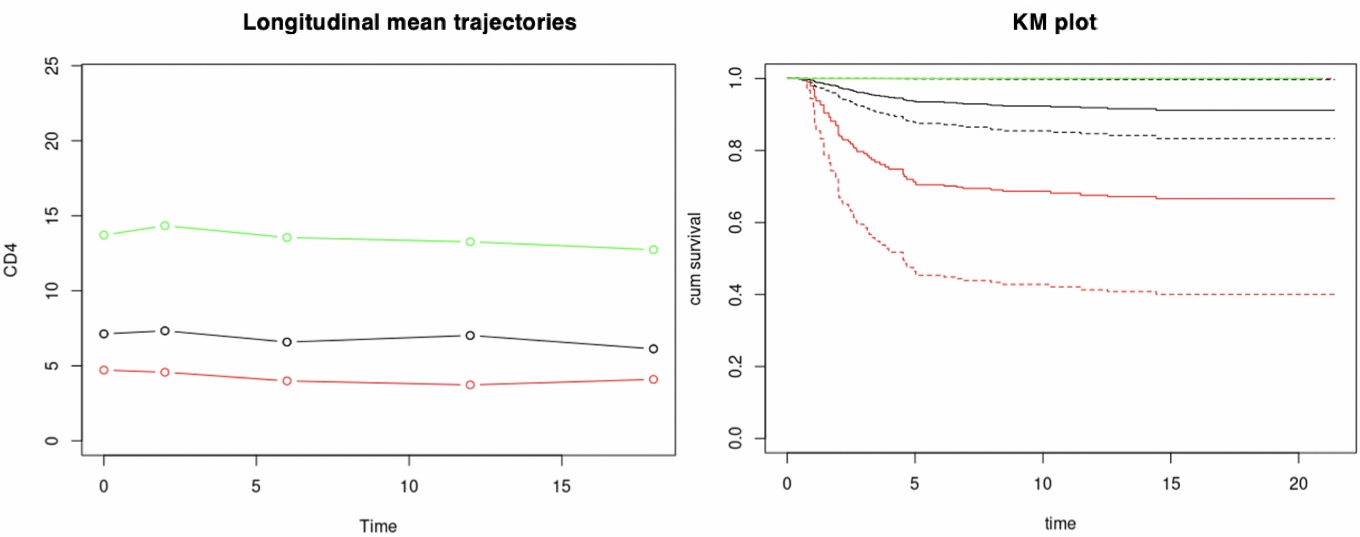}
					\end{center}
						\center{Fig. 5. Mean longitudinal trajectories without jumping behaviours (left) and K-M plot (black: no classification; red: class 2/low CD4 level; green:
						class 1/high CD4 level)}
				\end{figure}		

	\indent \quad	We take the first 2000 iterations as burn-in. The posterior means and sds, as well as 89\% credible intervals, for each parameter in the optimal model, are presented in Table  5. Variables $Time$, $PrevOI$ and intercept  provide  similar trends in the   longitudinal  model for both proposed and basic JLCM.	Compared with the results of the basic JLCM with optimal $K=3$ obtained using a frequentist approach via the lcmm package  and shown in Table B2 in Appendix B, we notice the CD4 level decreases with time.  $PrevOI$ has  similar effects as variable $Time$ in the longitudinal submodel  but  
	$PrevOI$ accounts for a sharper decrease in the  CD4 level. We also notice similar trends in terms of the effect of variables in the survival models compared to the basic JLCM.
	 In particular, both $PrevOI$  and $AZT$ can increase  risks in both  proposed and basic JLCM, however, $gender$ and $drug$ only seem to have a significant effect in our proposed JLCM.
	 The $\bm{\delta}_k$ values in the survival function are not zero, which indicates that there exists  association between the longitudinal and survival  processes. Non-zero $\bm{\xi}_{k2}$ values indicate there is evidence of a time-varying membership probability, providing further support for our proposed model. What is more, $\bm{\xi}_{12}$ is positive  while $\bm{\xi}_{22}$ is negative, which implies the membership probability of class  1 (2) increases  (decreases) with time.   \\ 
	 
	{ \centerline	{Table 5:	{Estimation results for the analysis of the aids data based on the time varying JLCM with $K$=2
			}
		}
	}	
	\setlength{\tabcolsep}{3.3mm}{
		\begin{longtable}{llllllcc}
			\hline
			\multicolumn{4}{c}{Class 1}	 
			&\multicolumn{4}{c}{Class 2} 
			\\ 
			\hline
			Parameter           & Estimate  &sd  & CI(89\% ETI)& 	Parameter  &Estimate &sd  & CI(89\% ETI)\\ \hline 
			$\beta_{11}$ &   15.0688      &        0.3179         &    [14.45, 15.70]	
			&	$\beta_{21}$ &  5.8386  & 0.2826&  [5.29, 6.38]\\
			$\beta_{12}$   &   -0.1214      &          0.0223       &    [-0.17, -0.08]	  
			&	$\beta_{22}$   
			&   -0.1134	  &  0.0146& [-0.14, -0.09]     	 \\
			$\beta_{13}$&     -0.1801   &         0.3327        &   [-0.82, 0.46]	&$\beta_{23}$  				      
			&   0.4044     &        0.2584         &   [-0.10, 0.90]	     \\
			$\beta_{14}$ &    -2.8503    &      0.3351          &     [-3.50, -2.20]&	$\beta_{24}$						 					 
			& -1.7465       &              0.1661   &      [-2.08, -1.42]	   \\
			$\beta_{15}$ &    -0.0999    &         0.3700        &   [-0.85, 0.61]	&	$\beta_{25}$ 											
			&      -0.2569  &           0.1486       &     [-0.55, 0.04]	   \\
			
			$\tau_1$    &     3.4486    &           0.3223      &     [2.79, 4.07]	 &$\tau_2$  				
			&     3.5991    &        0.3540         &    [2.86, 4.28]      	 \\   					  
			$\lambda_{1}$  &     0.0017    &       0.001        &     [0.0002, 0.0031]	&	$\lambda_2$    					  
			&      0.0105   &       0.0026           &    [0.01, 0.02]      	    \\				
			$	w_{11} $ &   -0.1396     &        0.0312         &      [-0.19, -0.08]&	$	w_{21} $  				
			&     -0.0613  &         0.0119       &    [-0.08, -0.04]	    \\				 
			$w_{12} $ &    1.0953    &       0.0969           &   [0.90, 1.28]	&$w_{22} $ 				 
			&    1.1993    &         0.1176     &     [0.95, 1.42]	    \\				 				 
			$	w_{13} $ &   0.2123      &     0.0314            &      [0.15, 0.27]	&$	w_{23} $  				 
			&   0.1824      &       0.0163          &    [0.11, 0.18] 	    \\					  
			$w_{14} $   &    0.1358     &          0.0126        &    [0.11, 0.16]	&	$w_{24} $  					  
			&     0.1824   &        0.0226          &     [0.14, 0.22] 	   \\			   
			$\delta_{11}$  &   0.1208     &            0.0312     &    [0.06, 0.18]	 &$\delta_{21} $  			   			   
			&     -0.0311    &          0.0078        &    [-0.04, -0.02]	  \\					    
			$\delta_{12} $&     0.1391   &    0.0357            &     [0.07, 0.20]&	$\delta_{22}$  					    
			&     0.1738   &    0.0447             &    [0.08, 0.25]		    \\				   
			$\xi_{11}$    &     -0.0453    &             0.0116     &      [-0.07, -0.02]	&	$\xi_{21}$   				   
			&     0.8022    &         0.0658       &    [0.66, 0.92]              \\ 
			$\xi_{12}$    &   0.0270     &         0.0053        &    [0.02, 0.04]&	$\xi_{22}$     				      
			&  -0.0130       &         0.0054        &    [-0.02,  0.00]	       	       \\					  				
			\hline
	\end{longtable}	}

		\indent \quad In table 6, we compare the posterior predictive probabilities for each group between the basic JLCM (under both our Bayesian approach and the standard likelihood based approach from the literature) and the proposed time varying JLCM. We notice that the majority of individuals are classified in class 2 (for the  time varying JLCM, this is the classification at the final time-point), with comparable percentages across the 3 models. Class 1 contains the second largest group of patients, with only very few in class 3  of the basic JLCMs. By allowing group membership to vary with time, our proposed model can capture the variability in the data without the need to introduce an additional class. \\
		
	{ \centerline{Table 6: 	{ Comparison of posterior membership probabilities between the  proposed and basic JLCM}
	}
}
	\begin{longtable}{llllllccc}
		\hline			
		Model      & Estimation method&$K$& \%class1 &\%class2 & \%class3 \\ 	\hline	
		basic JLCM & MLE	&3   & 19.27195 &73.44754  &7.280514   \\ 
		\hline	
		basic JLCM & Bayesian approach  &3  & 9.8501071&89.0792291& 1.0706638  \\ 
		\hline	
		proposed JLCM (Final group) & Bayesian approach &2 & 22.05567  &   77.94433&   \\ 
		\hline		
	\end{longtable}

		\subsection{Dynamic predictions of time to death}	
			\indent \quad 
		Similar to the simulation study, we are interested in using the CD4 measurements  to estimate the  expected
		survival probabilities and also assess how well we are able to discriminate between patients with high risk and low risk of dying. According to equations (9) and (10), we obtain  dynamic predictions based on our proposed JLCM for eight specific subjects, shown in Figures 6 and 7. We incorporate both the longitudinal trajectories and the survival probabilities on the same plot, to assess how various jumping behaviours might affect survival. The green dotted line represents the survival prediction based on the  proposed JLCM with 2 classes (the best model), the blue dotted line corresponds to the  survival  probabilities of the basic  JLCM with  3  classes (the best basic model) and the vertical brown dotted (purple dashed) lines indicate the time of event (censoring time). In the longitudinal trajectories, the black (red) circles indicate the subject is classified in class 1 (2) (based on the proposed JLCM) at the particular timepoint.\\
			\indent \quad According to Figure 6, subjects 2, 90 and 60, who exhibit no jumps, have a slowly decreasing survival probability around 80\% by the end of the study. For subject 60, the survival probability at the time of event based on the proposed JLCM (0.9341326
	) is more accurate compared to that based on the basic JLCM (0.9355598), although the estimation could be improved. For subject 188, who has a stable trajectory in class 2, with a large jump to class 1 at the last time point, the proposed JLCM seems to curb the decreasing survival probability slightly to account for the jump. \\
	\indent\quad 
In Figure 7, we also notice that our proposed model reacts faster to the steeply decreasing trajectory for subjects 68 and 318, adjusting the survival probability accordingly. Subject 75 also exhibits a jump to the lower CD4 group, however the change in the longitudinal response is minimal and thus the survival probability is higher in this instance (but still computed more accurately compared to the basic JLCM). The survival probability for subject 341 is also similar to subject 75, as both exhibit jumps between groups but minimal change in the longitudinal response. \\
	\indent\quad Compared with the basic JLCM, it is obvious that our proposed JLCM can provide more reasonable dynamic predictions as well as improve the accuracy of  predicted survival  probabilities. We can also see a link between the dynamic predictions from the proposed JLCM and the jumping trends exhibited in the longitudinal trajectories. The dynamic prediction plots of these eight subjects is shown in Figure C2 of Appendix C.\\
	
				\section{Conclusion and future work}
			\quad In this paper, we propose to extend the basic joint latent class model with a more general JLCM  which allows the class membership probability to change with  time. In our model, the longitudinal and survival processes are linked via both shared random effects and latent classes, and a Bayesian approach is used for estimation and inference. Through both simulation and real data analysis of the aids dataset, our proposed JLCM with time varying class membership probability, outperforms the basic JLCM, in terms of both DIC and estimation accuracy. In the dynamic prediction of survival probabilities, our model is generally more conservative than the basic JLCM and also more responsive to changes in the class membership over time. In the instances where the time to event is known, our model can predict survival probabilities more accurately, although there is still room for improvement. \\
			\indent \quad 
			In future  work,  the  variance-covariance structure of the random effects can also  be modelled simultaneously. For this particular joint model, rather than assuming the longitudinal and survival processes are joint via shared random effects, we   would consider different  random effects sharing the same distribution.
			It would then be useful to model the random effects covariance matrix because it might uncover any effects of covariates on the association between the joint models.
			This general joint latent class model  would contain  two more regression submodels for the heterogenous random covariance matrix,  which can be  decomposed using the Cholesky decomposition technique. To deal with outlying longitudinal measurements, a mixture distribution can also be considered for the longitudinal submodel or the normality assumption of  the random errors can be relaxed, by considering a skew-normal or $t$ distribution, leading to a more robust JLCM. 	\\ 
			
			\begin{figure}[H]
				\begin{center}
					\includegraphics[scale=0.68]{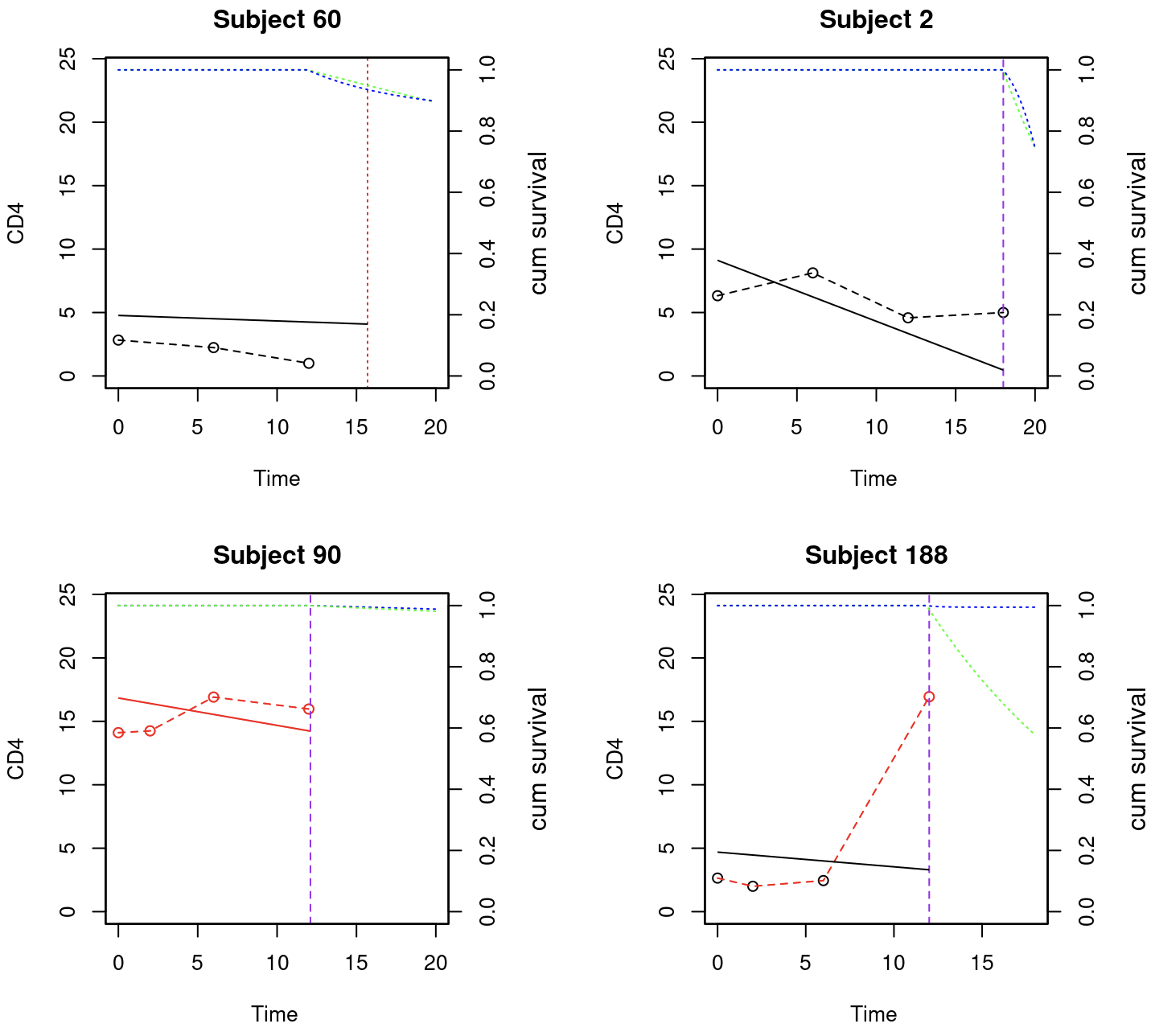}
				\end{center}
				\center{ Fig. 6 Prediction of aids data for four cases from both proposed and basic JLCM (black point: class 2/low CD4 level; red point: class 1/high CD4 level; black dashed  line: longitudinal trajectory which stayed in class 2; red dashed  line: longitudinal trajectory which stayed in class 1;  black full line: fitted trajectory which stayed in class 2; red  full line: fitted trajectory which stayed in class 1; 
					green dotted line: survival probability from proposed JLCM; blue dotted line: survival probability from basic JLCM; purple vertical dashed  line: censoring time; red vertical dotted line: time of event) 
				}
			\end{figure}
			
			\begin{figure}[H]
				\begin{center}
					\includegraphics[scale=0.69]{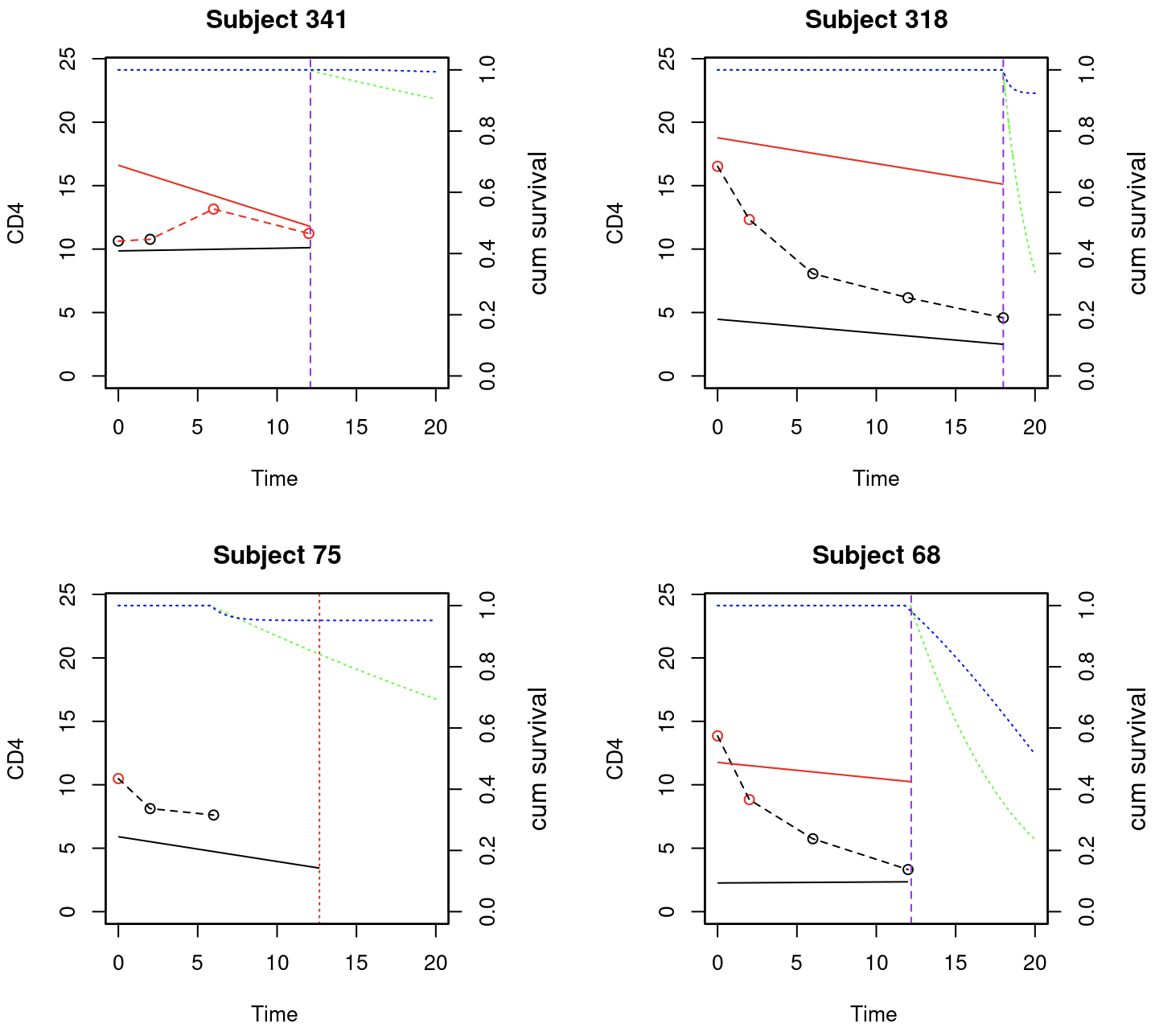}
				\end{center}
				\center{ Fig. 7 Prediction of aids data for four cases from both proposed and basic JLCM (black point: class 2/low CD4 level; red point: class 1/high CD4 level; black dashed  line: longitudinal trajectory which stayed in class 2; red dashed  line: longitudinal trajectory which stayed in class 1;  black full line: fitted trajectory which stayed in class 2; red  full line: fitted trajectory which stayed in class 1; 
					green dotted line: survival probability from proposed JLCM; blue dotted line: survival probability from basic JLCM; purple vertical dashed  line: censoring time; red vertical dotted line: time of event) 
				}
			\end{figure}

	\clearpage
	\appendix

	{\bf Appendix A: Full conditional densities}\\
	
	\textit{(1)Sampling $\bm{\beta}_{k}$ from normal distribution. }\\
	\indent \quad We set the prior distribution for $\bm{\beta}_{k}$ as $\pi(\bm{\beta}_{k}) \sim MVN(\bm{\beta}_{0}, \Sigma_{\beta_k})$ and then we can get: $\pi((\bm{\beta}_{k})	\varpropto exp(-\frac{1}{2}(\bm{\beta}_{k}-\bm{\beta}_{0})^T\Sigma_{\beta_k}^{-1}  (\bm{\beta}_{k}-\bm{\beta}_{0}))$.
	(Note: Denote $\pi(\cdot)$ as the density of the prior distribution and $\pi(\cdot|\cdot)$ as the density of the conditional distribution).\\
	Now,  we  get the conditional distribution for $\bm{\beta}_{k}$:
	\begin{align*}
		&\pi(\bm{\beta}_{k}|\cdot)\\
		\varpropto&\prod\limits_{i=1}^N\prod\limits_{j=1}^{m_i} \Bigg\{\exp\left[-\frac{(y_{ij}- \bm{X}_{2i}^T(t_{ij})\bm{\beta}_{k}-\bm{Z}_{i}^T(t_{ij})\bm{U}_i)^2}{2\tau_k}\right]\Bigg\}^{I(R_{ij}=k)}\cdot \pi(\bm{\beta}_{k})
		\\
		\varpropto&\exp\left[-\frac{\sum\limits_{i=1}^N\sum\limits_{j=1}^{m_i}2(y_{ij}-\bm{Z}_{i}^T(t_{ij})\bm{U}_i)(\bm{X}_{2i}^T(t_{ij})\bm{\beta}_{k}){I(R_{ij}=k)}-\sum\limits_{i=1}^N\sum\limits_{j=1}^{m_i}\bm{X}_{2i}^T(t_{ij})\bm{X}_{2i}(t_{ij})\bm{\beta}_{k}^T\bm{\beta}_{k}{I(R_{ij}=k)}}{2\tau_k}\right]\\
		&\exp(-\frac{1}{2}(\bm{\beta}_{k}-\bm{\beta}_{0})^T\Sigma_{\beta_k}^{-1}  (\bm{\beta}_{k}-\bm{\beta}_{0})) \\
		=&\exp\Bigg\{-\frac{1}{2}\Bigg\{\bm{\beta}_{k}^T(\frac{\sum\limits_{i=1}^N\sum\limits_{j=1}^{m_i}\bm{X}_{2i}^T(t_{ij})\bm{X}_{2i}(t_{ij}){I(R_{ij}=k)}}{\tau_k}+\Sigma_{\beta_k}^{-1})\bm{\beta}_{k}\\
		&-2(\bm{\beta}_{k}^T\frac{\sum\limits_{i=1}^N\sum\limits_{j=1}^{m_i}(y_{ij}-\bm{Z}_{i}^T(t_{ij})\bm{U}_i)\bm{X}_{2i}^T(t_{ij}){I(R_{ij}=k)}}{\tau_k}+\Sigma_{\beta_k}^{-1}\bm{\beta}_{0})\Bigg\}\Bigg\} 
	\end{align*}
	\quad If $\sum_{i=1}^N\sum_{j=1}^{m_i}I(R_{ij}=k)=0$, which means the likelihood makes no contributions to the posterior distribution, we sample $\bm{\beta}_{k}$ from the prior distribution $\pi(\bm{\beta}_{k})$. \\
	\indent \quad We know that if $y$ has the pdf $g(y)  \varpropto exp\{-{1}/{2}(Ay^2-2By)\}$, then we can say $y \sim N({B}/{A}, {1}/{A})$. And, for $\bm{\beta}_{k}$,  if $\sum_{i=1}^N\sum_{j=1}^{m_i}I(R_{ij}=k)\neq0$,
	$$\bm{\beta}_k|\cdot \sim N\left(\frac{B}{A}, \frac{1}{A}\right)$$
	Where $ A={\sum_{i=1}^N\sum_{j=1}^{m_i}\bm{X}_{2i}^T(t_{ij})\bm{X}_{2i}(t_{ij})I(R_{ij}=k)}/{\tau_k}+\Sigma_{\beta_k}^{-1}, B={\sum_{i=1}^N\sum_{j=1}^{m_i}(y_{ij}-\bm{Z}_{i}^T(t_{ij})\bm{U}_i)\bm{X}_{2i}^T(t_{ij})I(R_{ij}=k)}/{\tau_k}+\Sigma_{\beta_k}^{-1}\bm{\beta}_{0}$.\\
	\textit{(2)Sample $\lambda_{0k}$ from  Gamma distribution.}\\
	\indent \quad Specify the inverse Gamma prior for the gamma prior for each step of $kth$ baseline function $\lambda_{0k}$, leading to the conjugate posterior for $\lambda_{0k}$. i.e. $\pi(\lambda_{0k})$ follows Gamma distribution. And then we can get $\pi(\lambda_{0k}) \sim \Gamma(\alpha_{\lambda},\beta_{\lambda})$. 	
	Since $\lambda_{0k}(t)=\lambda_{0k}^{(s)}$ where $s=1,2, \dots S^k; t_{k}^{(s-1)}<t \le t_{k}^{(s)}; k=1,2, \dots K$, we can get the posterior distribution  as:
	\begin{align*}
		&\pi(\lambda_{0k}^{(s)}|\cdot)\\
		\varpropto &\Bigg\{\prod_{i=1}^{N}{\lambda_{0k}^{(s)}}^{I(T_i^*\le C_i)I(t_{k}^{(s-1)}<t \le t_{k}^{(s)})I(R_{ij}=k)}\\
		&exp\Bigg\{-I(t_{}\ge t_k^{(s-1)})I(R_{ij}=k)\lambda_{0k}^{(s)}\int_{t_k^{(s-1)}}^{min(t_{},t_{k}^{(s)})}\exp(\bm{X}_{3i}^T(u)\bm{\omega}_{k}+\bm{Z}_{i}^T(u_{})(\bm{\delta}_{k}\bm{U}_i))du\Bigg\}\Bigg\} \pi(\lambda_{0k}^{(s)})\\
		\varpropto&\Bigg\{\prod_{i=1}^{N}{\lambda_{0k}^{(s)}}^{I(T_i^*\le C_i)I(t_{k}^{(s-1)}<t \le t_{k}^{(s)})I(R_{ij}=k)}\\
		&exp\Bigg\{-I(t_{i}\ge t_k^{(s-1)})I(R_{ij}=k)\lambda_{0k}^{(s)}\int_{t_k^{(s-1)}}^{min(t_{},t_{k}^{(s)})}\exp(\bm{X}_{3i}^T(u)\bm{\omega}_{k}+\bm{Z}_{i}^T(u_{})(\bm{\delta}_{k}\bm{U}_i))du\Bigg\}\Bigg\} (\lambda_{0k}^{(s)})^{\alpha_{\lambda}-1}\exp(-{\beta_{\lambda}}{\lambda_{0k}^{(s)}}) \\		
		=&{\lambda_{0k}^{(s)}}^{\sum_{i=1}^{N}I(T_i^*\le C_i, t_{k}^{(s-1)}<t \le t_{k}^{(s)},R_{ij}=k)+ \alpha_{\lambda}-1}\\
		&\exp\Bigg\{-\lambda_{0k}^{(s)}(\sum_{i=1}^{N}I(t_{}\ge t_k^{(s-1)},R_{ij}=k)\int_{t_k^{(s-1)}}^{min(t_{},t_{k}^{(s)})}\exp(\bm{X}_{3i}^T(u)\bm{\omega}_{k}+\bm{Z}_{i}^T(u_{})(\bm{\delta}_{k}\bm{U}_i))du+\beta_{\lambda})\Bigg\}
	\end{align*}
	So we can see that if $\sum I(R_{ij}=k)\neq0$,\\
	$$\lambda_{0jk}^{(s)}|\cdot \sim \Gamma(E,F)$$
	where $$E=\alpha_{\lambda}+\sum_{i=1}^{N}I(T_i^*\le C_i, t_{k}^{(s-1)}<t \le t_{k}^{(s)}, R_{ij}=k)$$ $$F=\sum_{i=1}^{N}I(t_{}\ge t_k^{(s-1)}+\beta_{\lambda}, R_{ij}=k)\int_{t_k^{(s-1)}}^{min(t_{},t_{k}^{(s)})}\exp(\bm{X}_{3i}^T(u)\bm{\omega}_{k}+\bm{Z}_{i}^T(u_{})(\bm{\delta}_{k}\bm{U}_i))du+\beta_{\lambda}$$
	If $\sum I(R_{ij}=k)=0$, $\lambda_{0jk}^{(s)}|\cdot \sim \pi(\lambda_{0k})$.\\
\textit{	(3)Sampling $\Sigma_{ui}$ from  inverse Wishart distribution.}\\
	\indent \quad We use InverseWishart as the prior distribution for $\Sigma_{ui}$  i.e. $\pi(\Sigma_{ui}) \sim InverseWishart(\nu_u^*,{S_u^*}^{-1})$ (Schuurman et al., 2016), and then we can get the posterior distribution of $\Sigma_{i}$:
	$$\begin{aligned}
		&\pi((\Sigma_{ui}|\cdot)\\
		\varpropto & det(\Sigma_{ui})^{-\dfrac{1}{2}} exp\Big[-\dfrac{tr(\Sigma_{ui}^{-1}\bm{U}_i\bm{U}_i^T)}{2}\Big]\pi(\Sigma_{ui})\\
	\end{aligned}
	$$
	So we can see:
	$$\Sigma_{ui}|\cdot \sim InverseWishart(\nu_u^*+1,(\bm{U}_i\bm{U}_i^T+S_u^*)^{-1})$$\\
\textit{	(4)Sampling $\tau_k$ from inverse gamma distribution.}\\
	\indent \quad Further specify the inverse Gamma prior for the variance of the measurement error $\tau_k$, 
	i.e. we set the prior distribution for $\tau_k$: $\pi(\tau_k) \sim InverseGamma(\alpha^*,\beta^*) $ and then we can write: $\pi(\tau_k)={{\beta^*}^{\alpha^*}}/{\Gamma(\alpha^*)}(\tau_k)^{-\alpha^*-1}\exp(-{\beta^*}/{\tau_k})$ ($\alpha^*>0, \beta^*>0$).
	Thus, we can get the conditional distribution for $\tau_k$:
\begin{align*}
	&\pi(\tau_k|\cdot)\\
	\varpropto&\Bigg\{\prod_{i=1}^{N}\prod_{j=1}^{m_i}\left[(\tau_k)^{(-\frac{1}{2})}exp\left[-\frac{(y_{ij}- \bm{X}_{2i}^T(t_{ij})\bm{\beta}_{k}-\bm{Z}_{i}^T(t_{ij})U_i)^2}{2\tau_k}\right]\right]^{I(R_{i}=k)}\Bigg\}
	(\tau_k)^{-\alpha^*-1} exp(-\frac{\beta^*}{\tau_k})\\
	=&\Bigg\{\prod_{i=1}^{N}\prod_{j=1}^{m_i}(\tau_k)^{-\frac{I(R_{i}=k)}{2}}\Bigg\}(\tau_k)^{-\alpha^*-1}\exp\Bigg\{-\frac{\frac{\sum\limits_{i=1}^N\sum\limits_{j=1}^{m_i}(y_{ij}- \bm{X}_{2i}^T(t_{ij})\bm{\beta}_{k}-\bm{Z}_{i}^T(t_{ij})\bm{U}_i)^2}{2}I(R_{i}=k)+\beta^*}{\tau_k}\Bigg\}\\
	=&(\tau_k)^{-\frac{\sum_{i=1}^{N}m_iI(R_{i}=k)}{2}-\alpha^*-1}\exp\Bigg\{-\frac{\frac{\sum\limits_{i=1}^N\sum\limits_{j=1}^{m_i}(y_{ij}- \bm{X}_{2i}^T(t_{ij})\bm{\beta}_{k}-\bm{Z}_{i}^T(t_{ij})\bm{U}_i)^2I(R_{i}=k)}{2}+\beta^*}{\tau_k}\Bigg\}
\end{align*}
	
	So \begin{align*}
		\tau_k|\cdot \sim InverseGamma(&\frac{\sum_{i=1}^{N}m_iI(R_{i}=k)}{2}+\alpha^*,\\
		&\frac{\sum\limits_{i=1}^N\sum\limits_{j=1}^{m_i}(y_{ij}- \bm{X}_{2i}^T(t_{ij})\bm{\beta}_{k}-\bm{Z}_{i}^T(t_{ij})\bm{U}_i)^2I(R_{i}=k)}{2}+\beta^*)
	\end{align*}
	
	\indent \quad If $\sum_{i=1}^Nm_iI(R_{i}=k)=0$, we sample $\tau_k$ from $InverseGamma(\alpha^*,\beta^*)$; if not, we sample $\tau_k$ from\\ $InverseGamma({\sum_{i=1}^{N}m_iI(R_{i}=k)}/{2}+\alpha^*,
	{\sum_{i=1}^Nm_i(y_{ij}- \bm{X}_{2i}^T(t_{ij})\bm{\beta}_{k}-\bm{Z}_{i}^T(t_{ij})\bm{U}_i)^2I(R_{i}=k)}/{2}+\beta^*)$.\\
	
		{\bf Appendix B:  Estimation results for real data application on basic JLCM}\\
		
		\indent \quad We use basic JLCM with maximum likelihood  estimation  method to do the inference using lcmm package  in R. We set basic JLCM with group numbers from 1 to 4 and model it by the  function ``Jointlcmm". From Figure B1 in Appendix B, we can see mean trajectories as well as K-M plot of aids data for the optimal $K=3$ from basic JLCM in MLE method.  It is obvious that it makes sense to add time-varying membership probability into the basic JLCM for aids dataset. CD4 values are divided into different levels which indicates different health conditions for aids patients. The jumping behaviours can be considered for this specific aids data. Table B1 in Appendix B shows the summary information for basic JLCMs of aids data with various k settings. Aids data with 3 classes  gets the smallest DIC value among four model settings. Basic JLCM with three subgroups is  picked up as the  optimal model here. Estimation results of basic JLCM with $K=3$ are shown in Table B2.\\	
		
{ \centerline	{Table B1: 	{ Summary table of aids data for various k of basic JLCMs using lcmm}
}
}

	\begin{longtable}{llllllccc}
	\hline			
	Model      & loglik   & npm & BIC & \%class1 & \%class2 & \%class3 & \%class4\\ 	\hline	
$K$=1   & -4213.787   &15 & 8519.770 & 100.00000   \\ 
	\hline	
	$K$=2  & -4171.872 & 23 &8485.109 & 85.43897 &14.56103    \\ 
	\hline	
	$K$=3 & -4120.539&  31& 8431.613 & 19.27195 &73.44754  &7.280514   \\ 
	\hline	
	$K$=4  & -4122.770 & 39& 8485.248 & 30.19272 &14.56103 &40.899358& 14.3469\\ 
	\hline	
\end{longtable}

	{ \centerline	{Table B2:	{Estimation results for the analysis of the aids data based on the basic JLCM with optimal $K$=3 from lcmm package
		}
	}
}	
\setlength{\tabcolsep}{3.3mm}{
	\begin{longtable}{llllllccc}
		\hline
		\multicolumn{2}{c}{Class 1}	 
		&\multicolumn{2}{c}{Class 2} 
		&\multicolumn{2}{c}{Class 3} 
		\\ 
		\hline
		 & Estimate (sd)  &   &Estimate(sd)  & 	 &Estimate(sd) \\ \hline  
			$\beta_{11}$     \ \    \ \ &  14.8381(0.6838)	\ \ \ \ &	$\beta_{21}$   \ \    \ \ &   5.6112(1.2858)    \ \    \ \ &   $\beta_{31}$    \ \ \ \ &6.7059(0.3855)    \\		
		$\beta_{12}$         &   -0.1833(0.0294)&	$\beta_{22}$         &   -0.3262( 0.1646)     &   $\beta_{32}$  	&-0.1458(0.0195)   	 \\		
		$\beta_{13}$ &   0.2447(0.7148)       &$\beta_{23}$ &    0.7902(1.0377)       &  $\beta_{33}$  &0.0010(0.0347)	     \\				
		$\beta_{14}$ & -1.5153(0.7605)  	&	$\beta_{24}$ &   -3.2078(1.0278)     &    	$\beta_{34}$	 &-1.7904(0.4339) \\		
		$\beta_{15}$ &      -0.0777(0.9247)        	&	$\beta_{25}$& 0.3706(0.8545)     &   $\beta_{35}$	 &-0.3541(0.3152)   \\		
		$\sqrt{\tau}$          &    1.7478(0.0489)        &  $\sqrt{\tau}$        &    1.7478(0.0489)    & $\sqrt{\tau}$        &    1.7478(    0.0489)    	 \\   	
		$\lambda_1^*$		&0.3661(0.0278)&
		$\lambda_2^*$		&0.1530(0.0216)&
		$\lambda_3^*$		&0.0121(0.0204)\\
		$\nu_1^*$		& 1.4117(0.0838)&
		$\nu_2^*$		&1.2528(0.0939)&
		$\nu_3^*$		&0.4079(0.0804)\\		
		$	w_{1} $  &   0.1503(0.2504)          	&	$	w_{1} $  &   0.1503(0.2504)    	&	$	w_{1} $  &   0.1503( 0.2504)               \\		
		$w_{2} $  &    1.6962(0.3160)    	&$w_{2} $  &    1.6962(0.3160) &$w_{2} $  &    1.6962(0.3160) 		   \\
		$	w_{3} $   &   0.0402(0.1746)          &$	w_{3} $   &   0.0402(0.1746 )  &$	w_{3} $   &   0.0402(0.1746)        	    \\		
		$w_{4} $   &     -0.0536(0.1560)          &   	$w_{4} $   &     -0.0536(0.1560)	&   	$w_{4} $   &     -0.0536 (0.1560)     \\		
		$	\xi_{1}$   &  0.3339    (0.2505  )        &$		\xi_{2}$   &   0.2099      (0.2131)      &$	\xi_{3} $   &     0  ( /) \\
		\hline
\end{longtable}	}

	\begin{figure}[H]
	\begin{center}
		\includegraphics[scale=0.6]{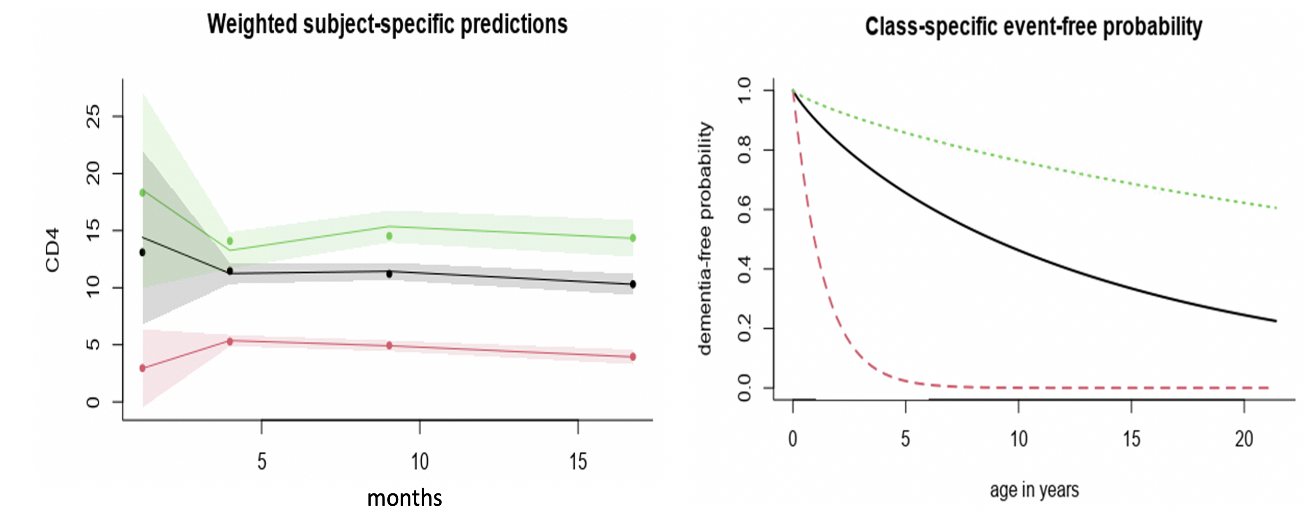}
			\center{Fig. B1.  Mean trajectories and  K-M plot of aids data from basic JLCMs using lcmm with $K=3$ (green: class 1/high CD4 level; red: class 2/low CD4 level; black: class 3/middle CD4 level)}
	\end{center}
\end{figure}

		{\bf Appendix C: Dynamic prediction plots of simulation and real data application}\\
		
			\begin{figure}[H]
			\begin{center}
				\includegraphics[scale=0.4]{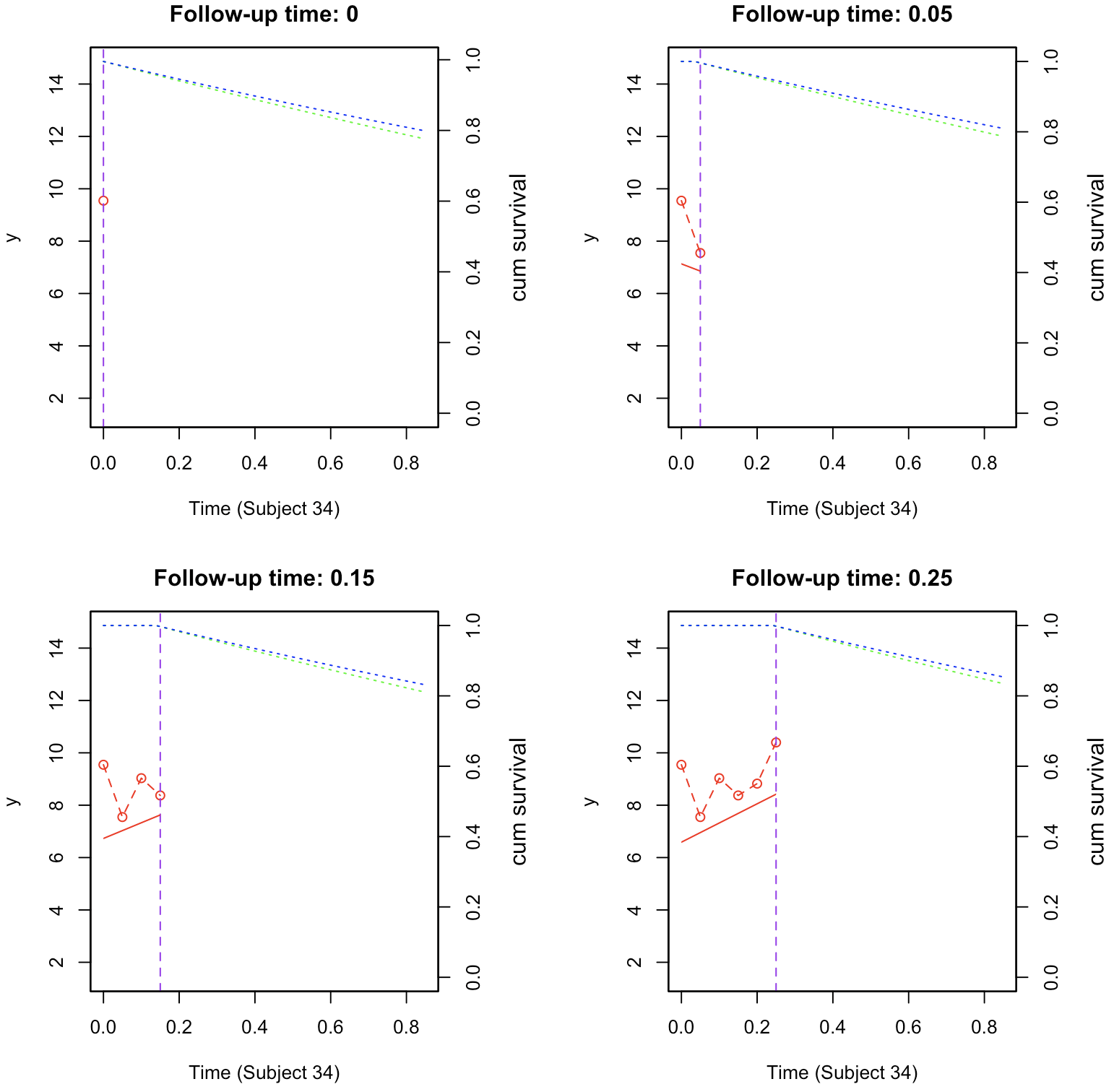}
			\end{center}
		\end{figure}
		\begin{figure}[H]
			\begin{center}
				\includegraphics[scale=0.4]{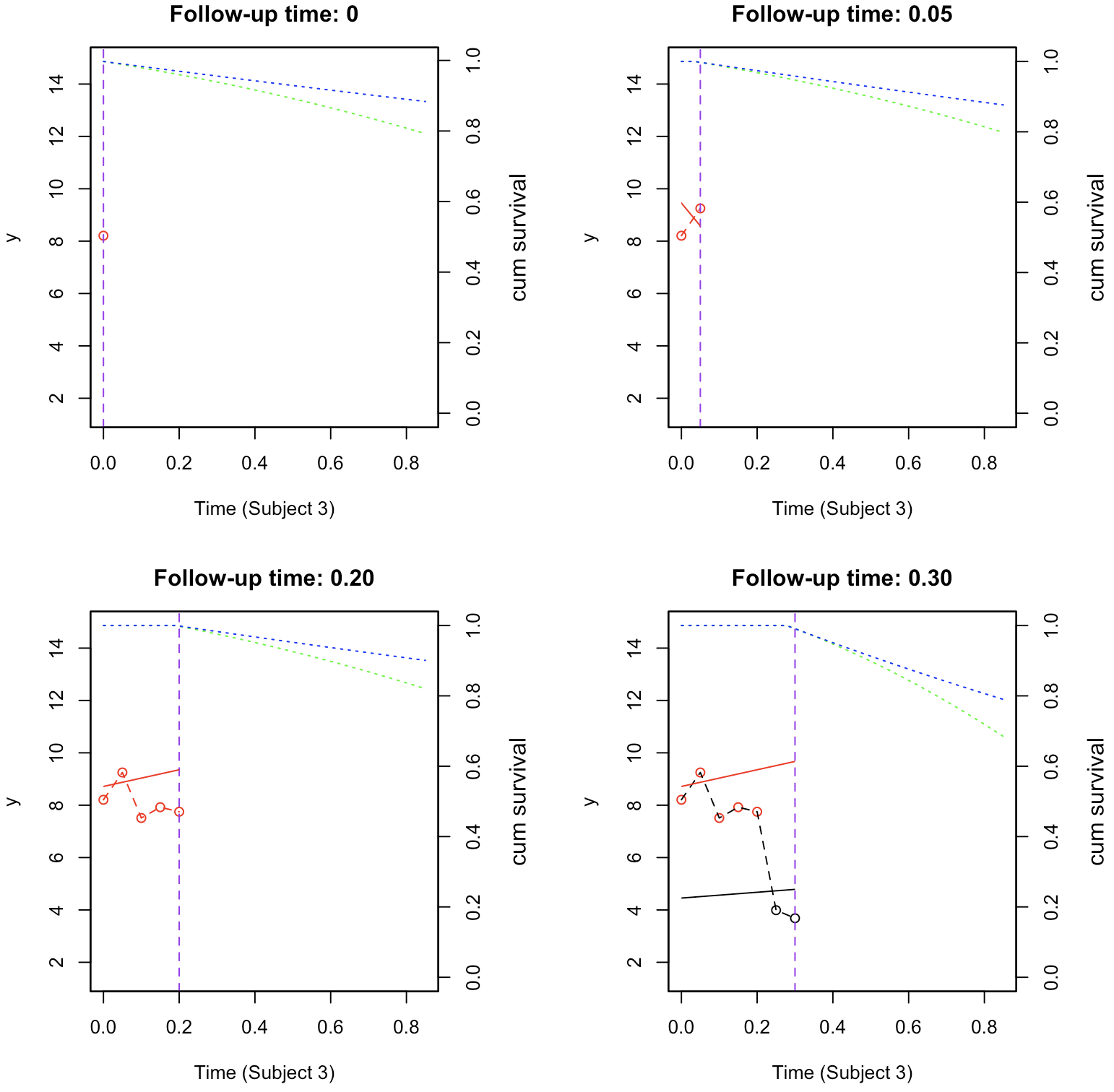}
			\end{center}
		\end{figure}
		\begin{figure}[H]
			\begin{center}
				\includegraphics[scale=0.4]{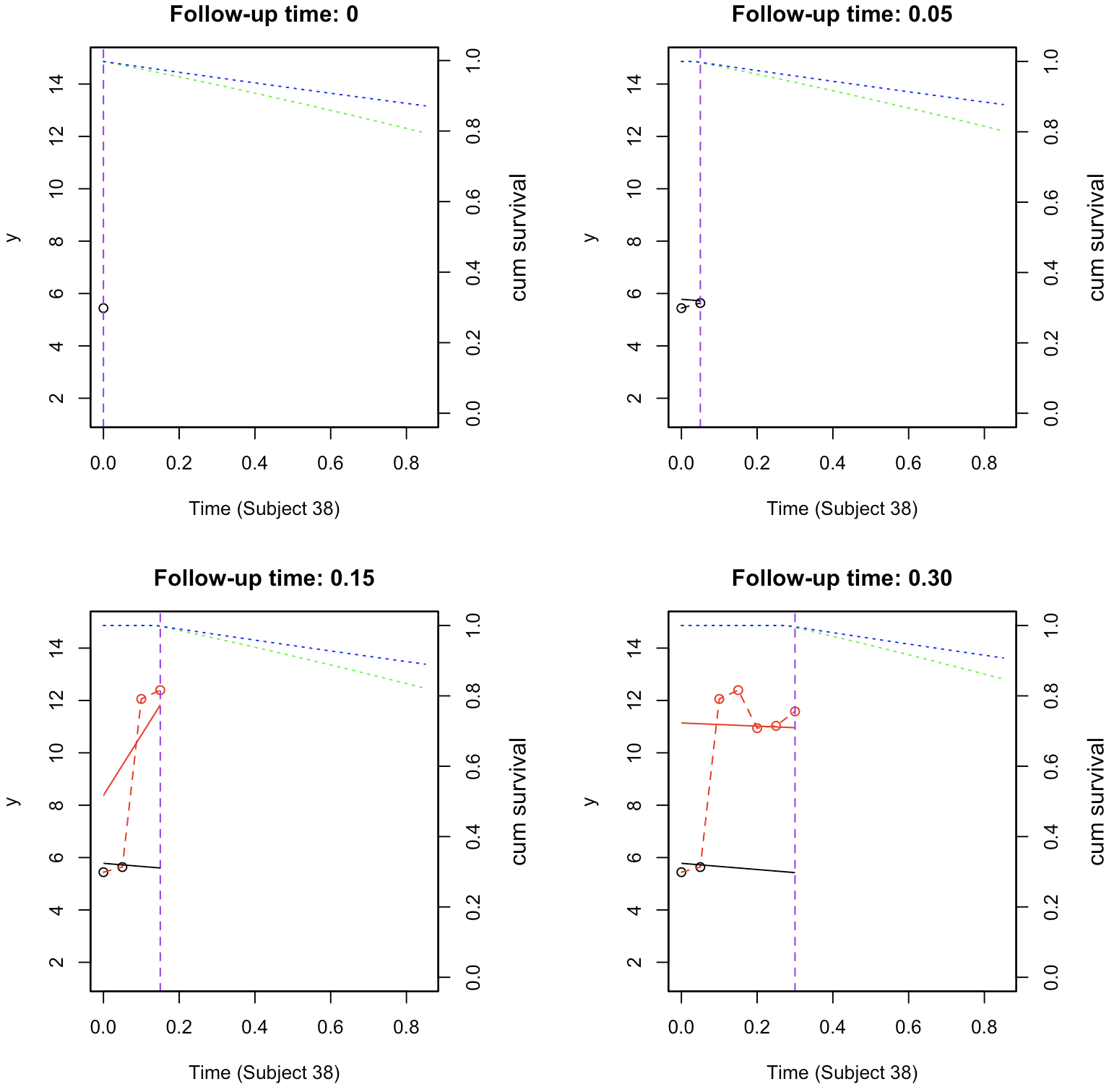}
			\end{center}
		\end{figure}
		\begin{figure}[H]
			\begin{center}
				\includegraphics[scale=0.4]{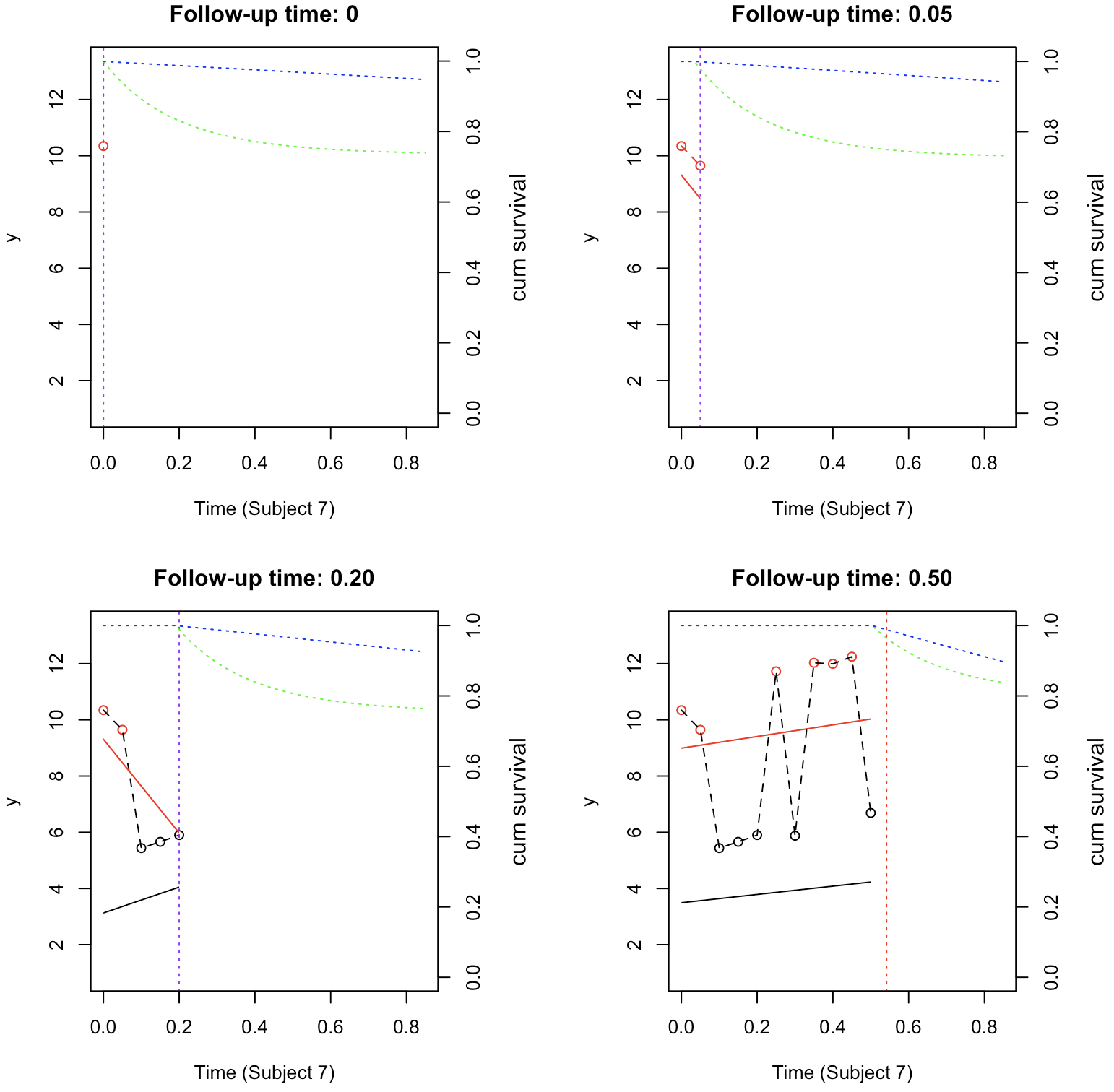}
			\end{center}
		\end{figure}
		\begin{figure}[H]
			\begin{center}
				\includegraphics[scale=0.35]{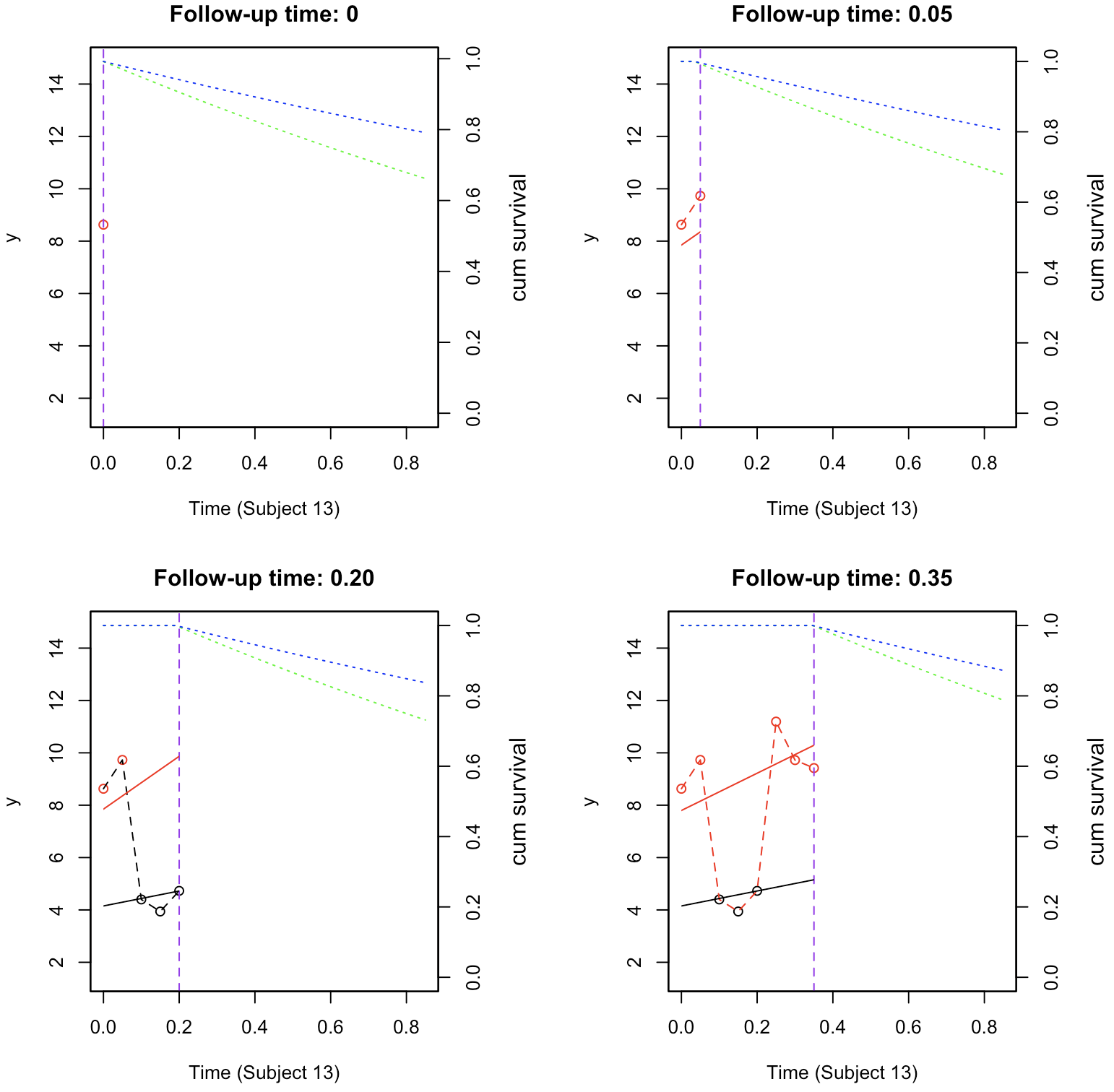}
			\end{center}
		\end{figure}
		\begin{figure}[H]
				\begin{center}
		\includegraphics[scale=0.35]{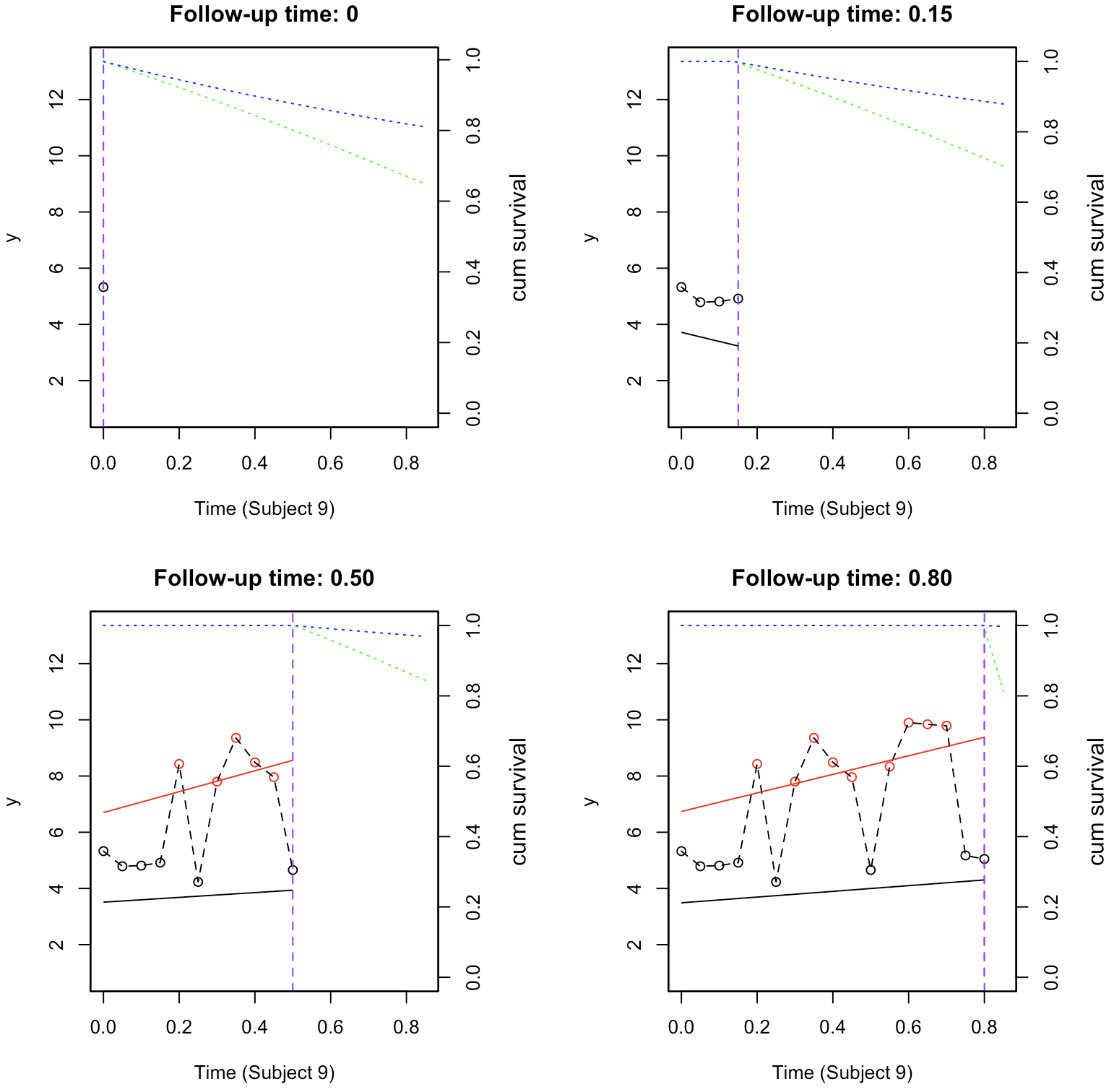}
		\center{ Fig. C1. Dynamic prediction of simulated data for six cases from both proposed and basic JLCM (black point: class 1/low level; red point: class 2/high level; black dashed  line: longitudinal trajectory which stayed in class 1; red  dashed  line: longitudinal trajectory which stayed in class 2; black full line: fitted trajectory which stayed in class 1; red  full line: fitted trajectory which stayed in class 2;  green dotted line: survival probability from proposed JLCM; blue dotted line: survival probability from basic JLCM; purple vertical dashed line: subject censored; red vertical  dotted line: time to event happened) }
			\end{center}
	\end{figure}

	\begin{figure}[H]
		\begin{center}
			\includegraphics[scale=0.5]{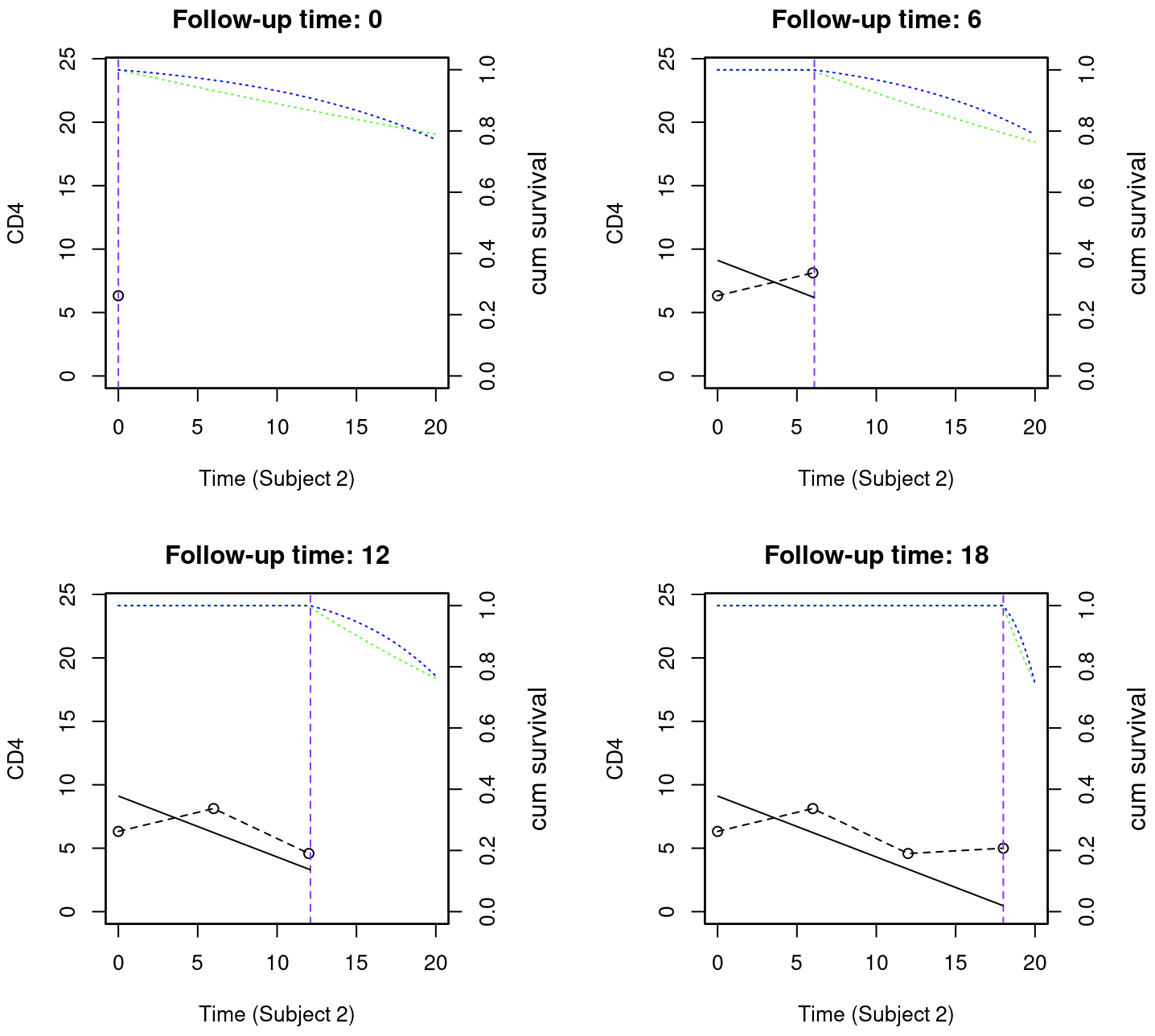}
		\end{center}
	\end{figure}

\begin{figure}[H]
	\begin{center}
		\includegraphics[scale=0.5]{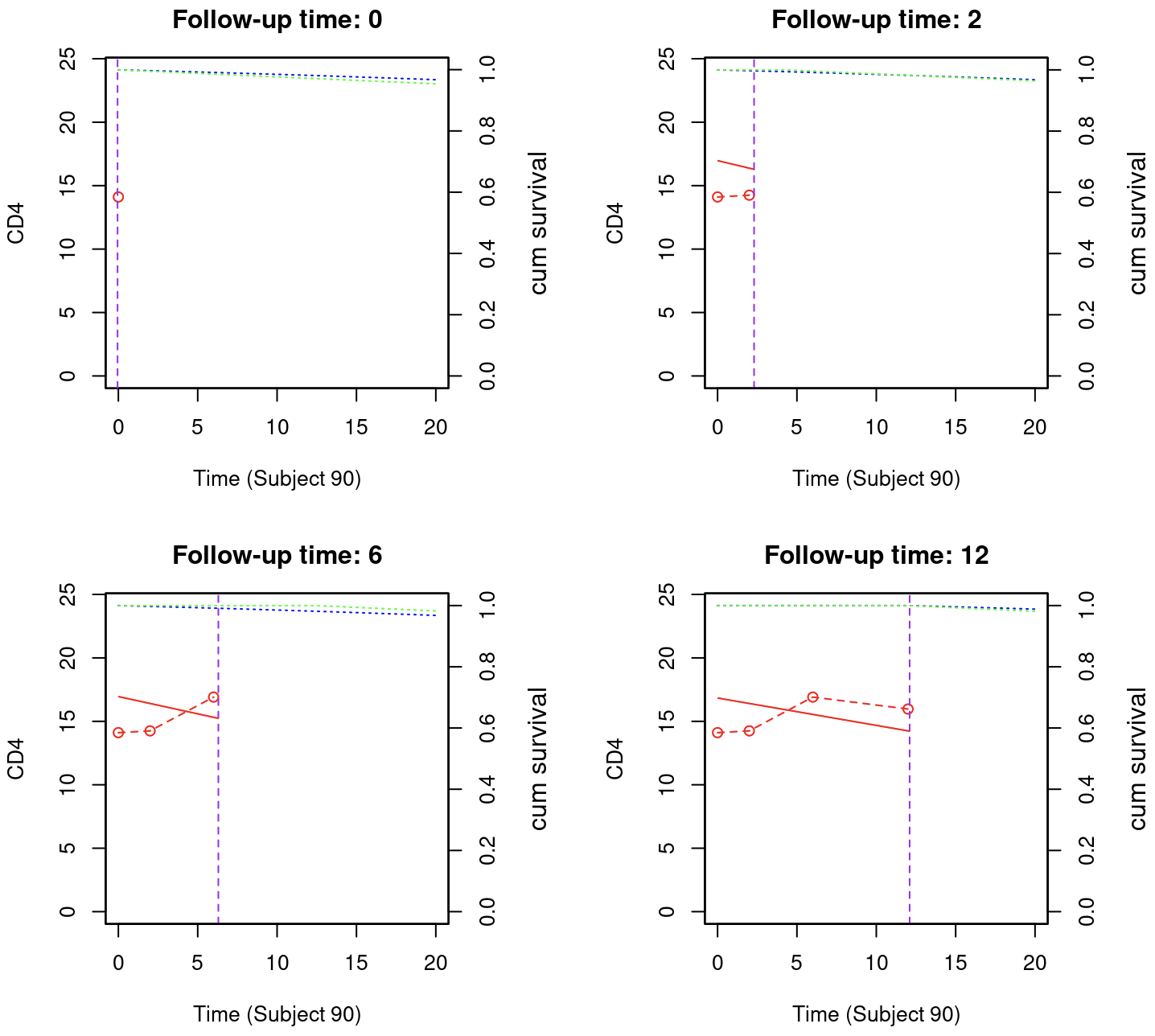}
	\end{center}
\end{figure}

	\begin{figure}[H]
	\begin{center}
		\includegraphics[scale=0.5]{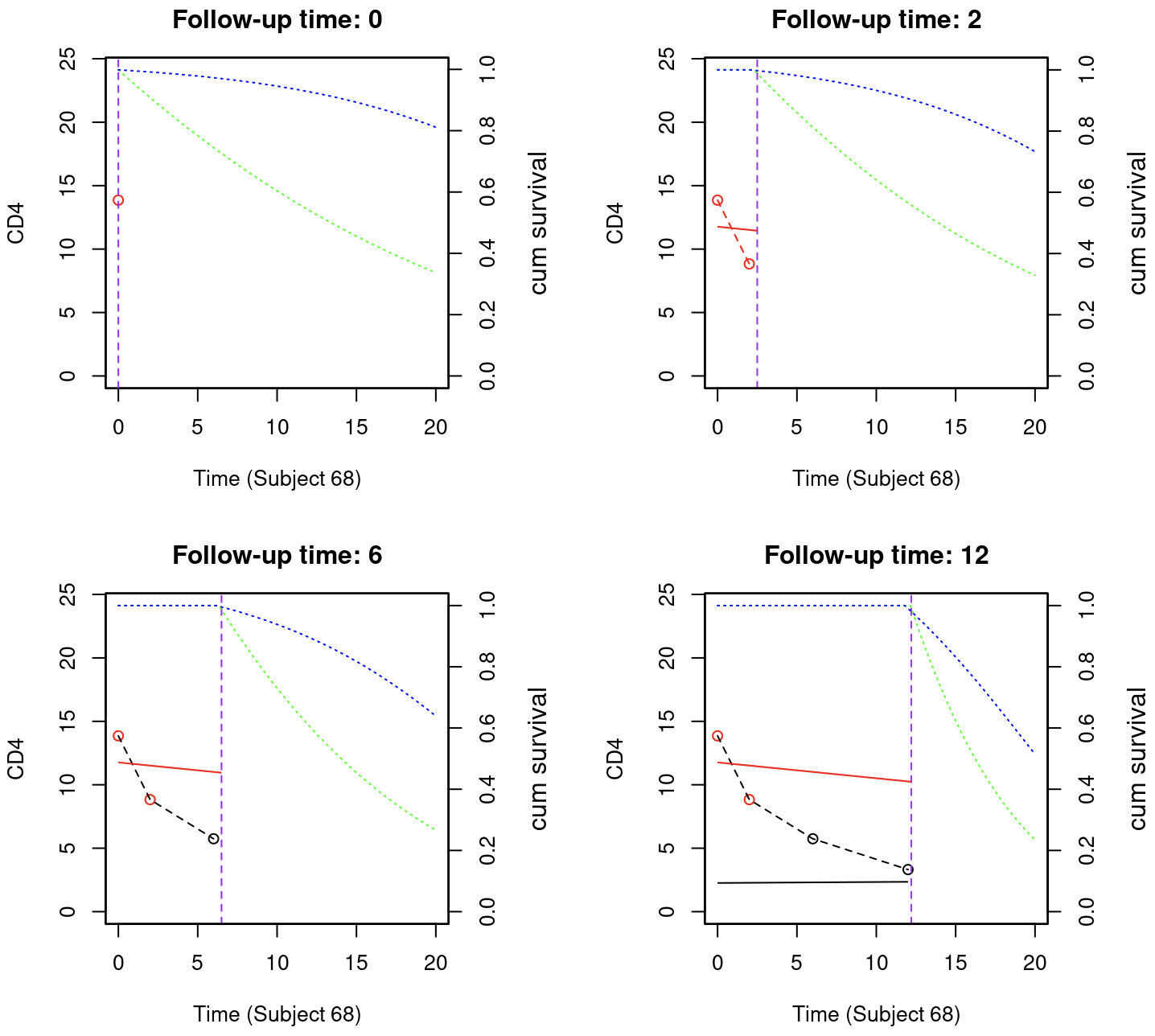}
	\end{center}
\end{figure}

	\begin{figure}[H]
	\begin{center}
		\includegraphics[scale=0.5]{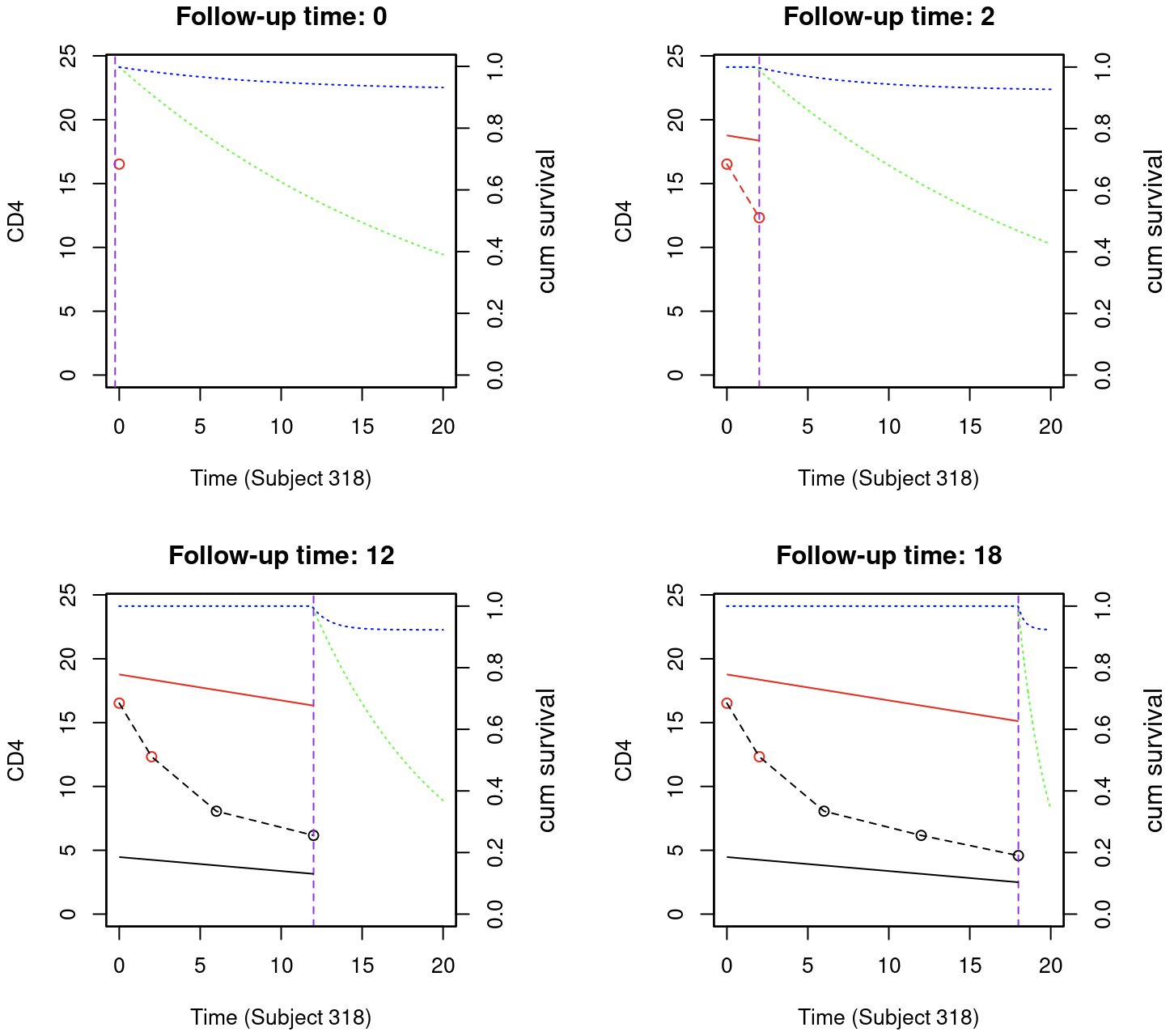}
	\end{center}
\end{figure}

		\begin{figure}[H]
		\begin{center}
			\includegraphics[scale=0.5]{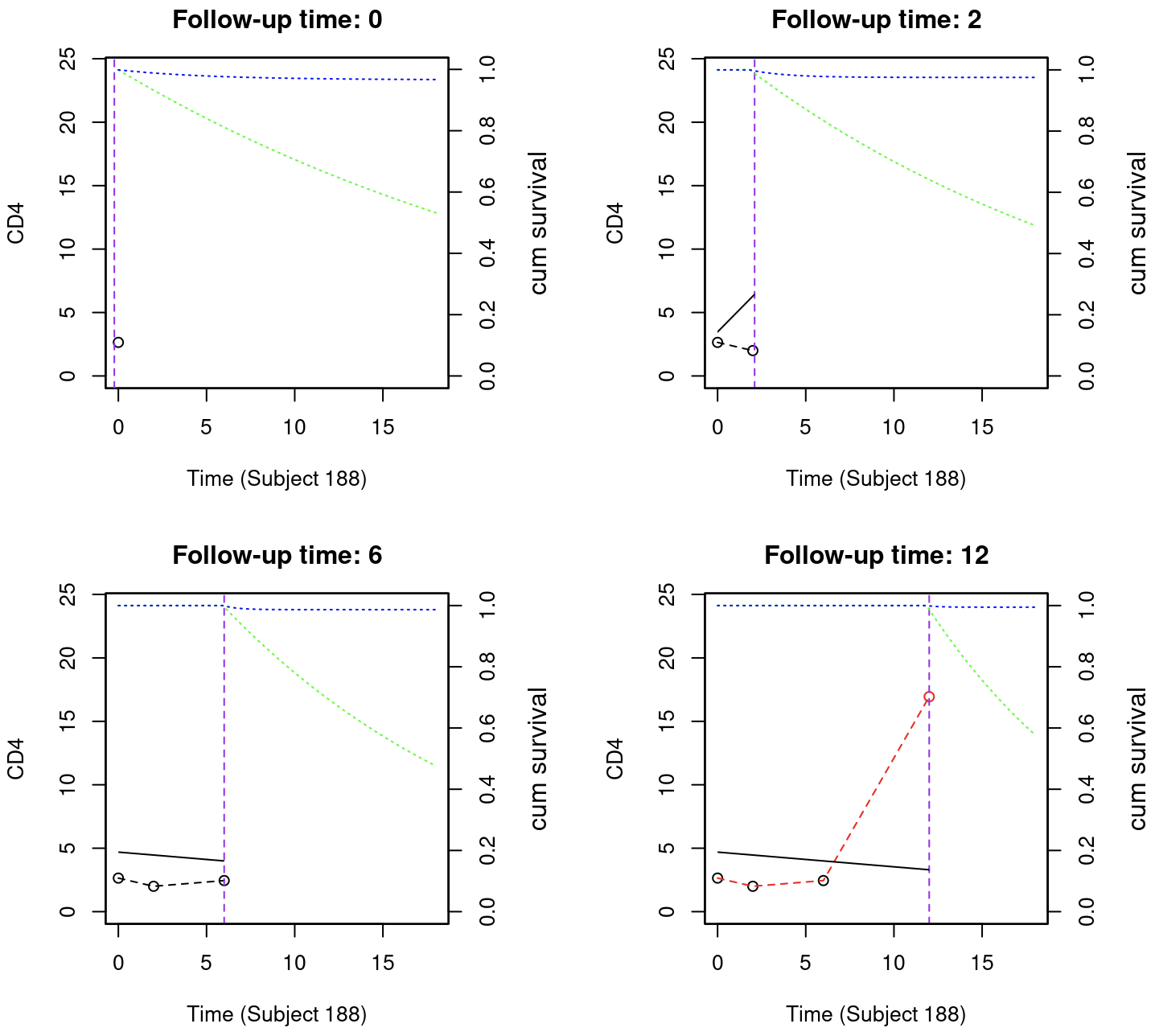}
		\end{center}
	\end{figure}

	\begin{figure}[H]
	\begin{center}
		\includegraphics[scale=0.5]{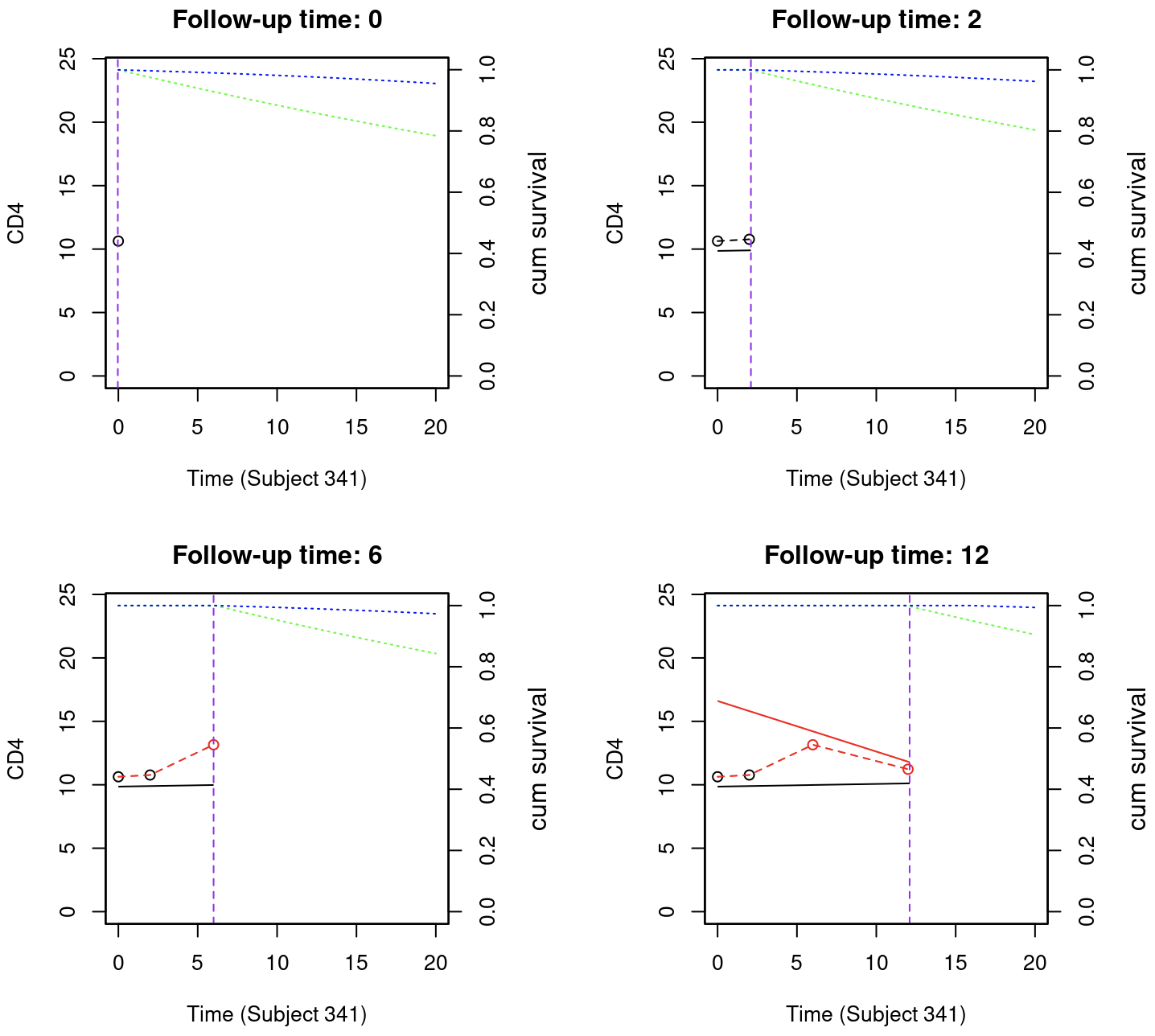}
	\end{center}
\end{figure}

	\begin{figure}[H]
	\begin{center}
		\includegraphics[scale=0.45]{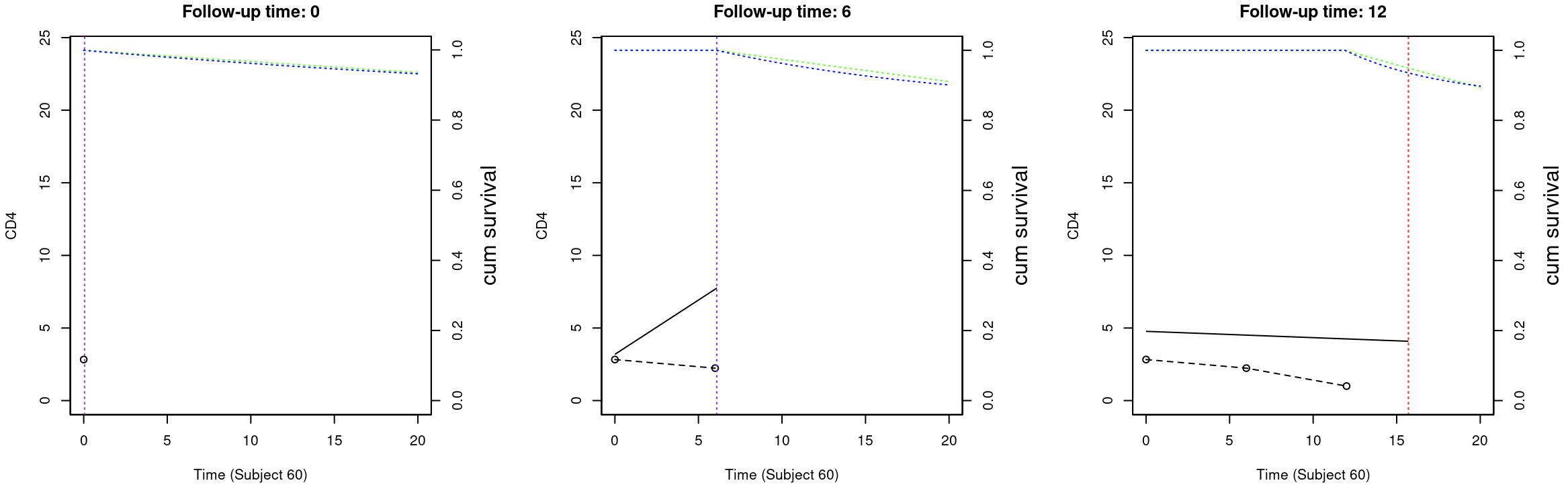}
	\end{center}
\end{figure}

	\begin{figure}[H]
			\begin{center}
		\includegraphics[scale=0.45]{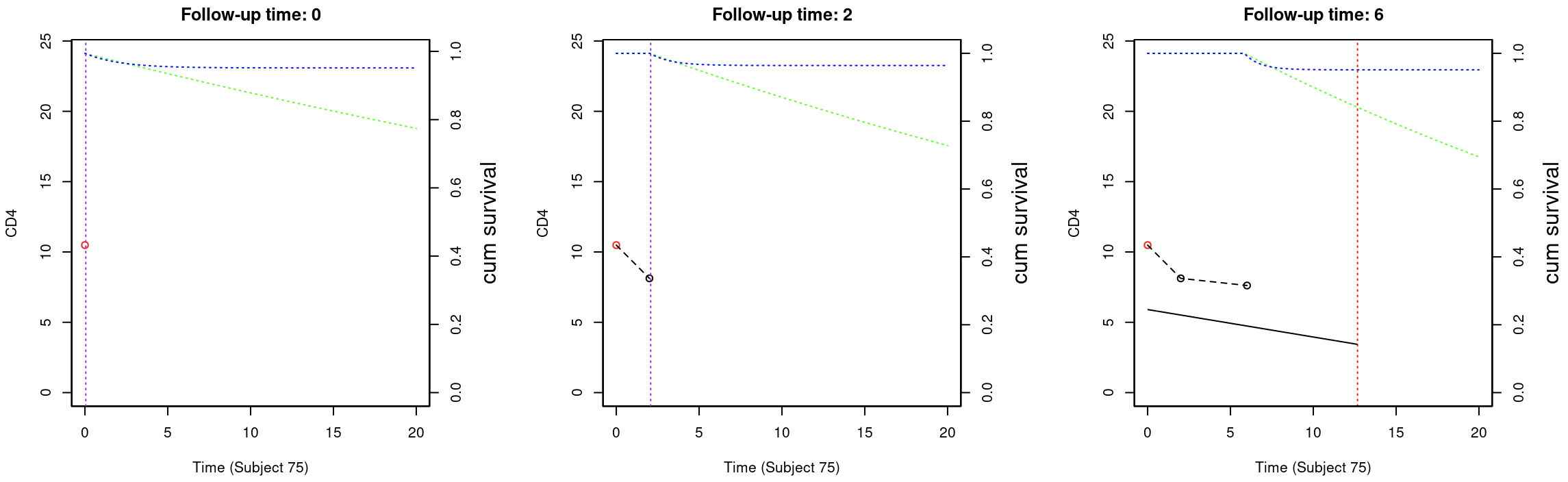}
		\center{ Fig. C2. Dynamic prediction of aids data for eight cases from both proposed and basic JLCM (black point: class 2/low CD4 level; red point: class 1/high CD4 level; black dashed  line: longitudinal trajectory which stayed in class 2; red dashed  line: longitudinal trajectory which stayed in class 1;  black full line: fitted trajectory which stayed in class 2; red  full line: fitted trajectory which stayed in class 1; 
			green dotted line: survival probability from proposed JLCM; blue dotted line: survival probability from basic JLCM; purple vertical dashed  line: censoring time; red vertical dotted line: time of event) }
			\end{center}
	\end{figure}


\begin{thebibliography}{999}
			\bibitem{ANO}
		Aninopoulou, E., Nasserinejad, K., Szczesniak , R., and Rizopoulos, D. (2020). Integrating latent classes in the Bayesian shared parameter joint model of longitudinal and survival outcomes. Statistical methods in medical research, 29(11), pp.3294–3307.
	
				\bibitem{ANO}
		Bartolucci, F. and Farcomeni, A. (2019). A shared‐parameter continuous‐time hidden Markov and survival model for longitudinal data with informative dropout. Statistics in medicine, 38(6), pp.1056–1073.
			\bibitem{ANO}
		Crowther, M.J., Abrams, K.R. and Lambert, P.C. (2013). Joint Modeling of Longitudinal and Survival Data. The Stata journal, 13(1), pp.165–184.
		
			\bibitem{ANO}
		Chiang, C.-T. (2009). A more flexible joint latent model for longitudinal and survival time data. Metrika, 73(2), pp.151–170.
		\bibitem{ANO}
	Dai, H. and Pan, J. (2018). Joint modelling of survival and longitudinal data with informative observation times.
	
	
		
			\bibitem{ANO} 
		De Gruttola, V. and Tu, X.M. (1994). Modelling Progression of CD4-Lymphocyte Count and Its Relationship to Survival Time. Biometrics, 50(4), pp.1003–1014.
	
		\bibitem{ANO}
	Elashoff, R.M., Li, G. and Li, N. (2008). A Joint Model for Longitudinal Measurements and Survival Data in the Presence of Multiple Failure Types. Biometrics, 64(3), pp.762–771.
	
	
		\bibitem{ANO} 
		Garre, F.G., Zwinderman, A.H., Geskus, R.B., and Sijpkens. Y.W.J. (2008). A joint latent class changepoint model to improve the prediction of time to graft failure. Journal of the Royal Statistical Society. Series A, Statistics in society, 171(1), pp.299–308.
		
				\bibitem{ANO} 
		Gelman, A., Gilks, W.R. and Roberts, G.O. (1997). Weak Convergence and Optimal Scaling of Random Walk Metropolis Algorithms. The Annals of applied probability, 7(1), pp.110–120.
	
			\bibitem{ANO} 
	Goldman, A.I., Carlin, B.P., Crane, L.R., Launer, C., Korvick, J.A., Deyton,L., and Abrams, D.I. (1996). Response of CD4 lymphocytes and clinical consequences of treatment using ddI or ddC in patients with advanced HIV infection. Journal of acquired immune deficiency syndromes and human retrovirology, 11(2), pp.161–169.
	
	
				\bibitem{ANO} 
	Hougaard, P. (1995). Frailty models for survival data. Lifetime data analysis, 1(3), pp.255–273.
	
			\bibitem{ANO} 
	Hu, W., Li, G. and Li, N. (2009). A Bayesian approach to joint analysis of longitudinal measurements and competing risks failure time data. Statistics in medicine, 28(11), pp.1601–1619.
	
			\bibitem{ANO} 
Haario, H., Saksman, E. and Tamminen, J. (2001). An Adaptive Metropolis Algorithm. Bernoulli: official journal of the Bernoulli Society for Mathematical Statistics and Probability, 7(2), pp.223–242.
	\bibitem{ANO}
Hastings, W. K. (1970). Monte carlo sampling methods using markov chains and their applications.

		\bibitem{ANO}
		Henderson, R., Diggle, P. and Dobson, A. (2000). Joint modelling of longitudinal measurements and event time data. Biostatistics (Oxford, England), 1(4), pp.465–480.

		\bibitem{ANO} 
	Huang, X., Li, G., Elasho, R.M., and Pan, J. (2011). A general joint model for longitudinal measurements and competing risks survival data with heterogeneous random effects.
	
		\bibitem{ANO} 
		Ibrahim, J.G., CHU, H.T. and CHEN, L.M. (2010). Basic Concepts and Methods for Joint Models of Longitudinal and Survival Data. Journal of clinical oncology, 28(16), pp.2796–2801.
		
		\bibitem{ANO} 
	Kürüm, E., Jeske, D.R., Behrendt, C.E., and Lee, P. (2018). A copula model for joint modeling of longitudinal and time‐invariant mixed outcomes. Statistics in medicine, 37(27), pp.3931–3943.

\bibitem{ANO} 
Kruschke, J. (2014). Doing Bayesian Data Analysis: A Tutorial with R, JAGS, and Stan. Academic Press.

		\bibitem{ANO} 
	Lin, H., Turnbull, B.W., McCulloch, C.E., and Slate, E.H. (2002). Latent Class Models for Joint Analysis of Longitudinal Biomarker and Event Process Data: Application to Longitudinal Prostate-Specific Antigen Readings and Prostate Cancer. Journal of the American Statistical Association, 97(457), pp.53–65.
			\bibitem{ANO} 
	Lin, H., Han, L., Peduzzi, P.N., Murohy, T.E., Gill, T.M., and Allore, H.G.(2014). A dynamic trajectory class model for intensive longitudinal categoricaloutcome. Statistics in medicine, 33(15).
	
		\bibitem{ANO} 
Liu, Y., Liu, L. and Zhou, J. (2015). Joint latent class model of survival and longitudinal data: An application to CPCRA study. Computational statistics \& data analysis, 91, pp.40–50.

	\bibitem{ANO} 
Metropolis, N., Rosenbluth A.W., Rosenbluth, M.N., Teller A.H., and Teller,E.  (1953). Equation of State Calculations by Fast Computing Machines. The Journal of chemical physics, 21(6), pp.1087–1092.

\bibitem{ANO}
Makowski, D., Ben-Shachar, M. S., and Lüdecke, D. (2019). BayestestR: Describing Effects and Their Uncertainty, Existence and Significance within the Bayesian Framework. Journal of Open Source Software, 4(40), 1541. https://doi.org/10.21105/joss.01541

	\bibitem{ANO}
McElreath, R. (2018). Statistical Rethinking: A Bayesian Course with Examples in R and Stan. Chapman; Hall/CRC.

	\bibitem{ANO} 
Murtaugh, P.A., Dickson, E., Vandam, G., Malinchoc, M., Grambsch, P., Langworthy, A., and Gips, C. (1994). Primary biliary cirrhosis: prediction of short- term survival based on repeated patient visits. Hepatology (Baltimore, Md.), 20(1), pp.126–134.


			\bibitem{ANO} 
		Philipson, P. et al.  (2012). JoineR: Joint modelling of repeated measurements and time-to-event data. Comprehensive R Archive Network, United Kingdom. http://cran.r-project.org/web/packages/joineR/index.html 
	
		\bibitem{ANO} 
	Proust-Lima, C., Dartigues, J.F., and Jacqmin-Gadda, H.  (2014). Joint latent class models for longitudinal and time-to-event data: A review. Statistical methods in medical research, 23(1), pp.74–90.
	
		\bibitem{ANO} 
	Proust-Lima, C., Dartigues, J.-F. and Jacqmin-Gadda, H. (2016). Joint modeling of repeated multivariate cognitive measures and competing risks of dementia and death: a latent process and latent class approach. Statistics in medicine, 35(3), pp.382–398.
	
		\bibitem{ANO} 
	Proust-Lima, C., Philipps, V. and Liquet, B. (2017). Estimation of Extended Mixed Models Using Latent Classes and Latent Processes: The R Package lcmm. Journal of statistical software, 78(2), pp.1–56.
	
		\bibitem{ANO} 
Rizopoulos, D., Verbeke, G. and Lesaffre, E. (2009). Fully exponential Laplace approximations for the joint modelling of survival and longitudinal data. Journal of the Royal Statistical Society. Series B, Statistical methodology, 71(3), pp.637–654.

		\bibitem{ANO}
	Roberts, G.O. and Rosenthal, J.S. (2009). Examples of Adaptive MCMC. Journal of computational and graphical statistics, 18(2), pp.349–367.
	
		\bibitem{ANO}
	Roberts, G.O. and Rosenthal, J.S. (2001). Optimal Scaling for Various Metropolis-Hastings Algorithms. Statistical science, 16(4), pp.351–367.
	
			\bibitem{ANO} 
	Rizopoulos, D. (2010). JM: An R package for the joint modelling of longitudinal and time-to-event data. Journal of statistical software, 35(9), pp.1–33.
	
			\bibitem{ANO}
			Rizopoulos D. (2016). The R Package JMbayes for Fitting Joint Models for Longitudinal and Time-to-Event Data Using MCMC. Journal of Statistical Software, 72(1), 1–46. 
			
				\bibitem{ANO} 
		Rouanet, A. , Joly, P., Dartigues, J.F., Proust-Lima, C., and Jacqmin-Gadda, H.  (2016). Joint latent class model for longitudinal data and interval-censored semi-competing events: Application to dementia. Biometrics, 72(4), pp.1123–1135.
		
						\bibitem{ANO}
	Rizopoulos, D. (2011). Dynamic Predictions and Prospective Accuracy in Joint Models for Longitudinal and Time-to-Event Data. Biometrics, 67(3), pp.819–829.
	
	
		\bibitem{ANO} 
		Schluchter, M.D. (1992). Methods for the analysis of informatively censored longitudinal data. Statistics in medicine, 11(14-15), pp.1861–1870.
		
		\bibitem{ANO} 
		Self, S., Pawitan, Y. (1992). Modeling a Marker of Disease Progression and Onset
		of Disease. In: Jewell N.P., Dietz K., Farewell V.T. (eds) AIDS Epidemiology. MA.DOI:https://doi.org/10.1007/978-1-4757-1229-2-11
		
		\bibitem{ANO} 
	Schuurman, N.K., Grasman, R.P.P.P. and Hamaker, E.L. (2016). A Comparison of Inverse-Wishart Prior Specifications for Covariance Matrices in Multilevel Autoregressive Models. Multivariate behavioral research, 51(2-3), pp.185–206.
	
		\bibitem{ANO} 
	Troxel, A.B., Harrington, D.P. and Lipsitz, S.R. (1998). Analysis of longitudinal data with non-ignorable non-monotone missing values. Journal of the Royal Statistical Society: Series C (Applied Statistics), 47(3), pp.425–438.
	\bibitem{ANO} 
	Tsiatis, A. A., Degruttola, V. and Wulfsohn, M.S. (1995). Modeling the relationship of survival to longitudinal data measured
	with error. applications to survival and cd4 counts in patients with aids. Journal of the American Statistical Association  90, 27--37.


	\bibitem{ANO} 
	Wang, C. Y.  (2006). CORRECTED SCORE ESTIMATOR FOR JOINT MODELING OF LONGITUDINAL AND FAILURE TIME DATA. Statistica Sinica, 16(1), 235–253.

			\bibitem{ANO}
	Zhou, X., Kang, K., Kwok, T., and Song, X.  (2021) Joint Hidden Markov Model for Longitudinal and Time-to-Event Data with Latent Variables, Multivariate Behavioral Research, DOI: 10.1080/00273171.2020.1865864
				\bibitem{ANO}
		Zhang, N., and Simono, J.S. (2020). The Potential for Nonparametric Joint Latent Class Modeling of Longitudinal and Time-to-Event Data. In: La Rocca, M., Liseo, B., Salmaso, L. (eds) Nonparametric Statistics. ISNPS 2018. SpringerProceedings in Mathematics \& Statistics, vol 339. Springer, Cham.
			\bibitem{ANO}
			Zhang, N. and Simono, J.S. (2022). Joint latent class trees: A tree-based approachto modeling time-to-event and longitudinal data. Statistical methods inmedical research, 31(4), pp.719-752.
			\bibitem{ANO}
	Zhang, Z., Charalambous, C. and Foster, P. (2021). A Gaussian copula joint model for longitudinal and time-to-event data with random effects.
	\end{thebibliography}
\end{document}